\documentclass[letterpaper,twocolumn,10pt]{article}
\usepackage{usenix2019_v3}
\usepackage[english]{babel}
\usepackage{amsmath,amssymb,amsfonts}
\usepackage{graphicx}
\usepackage{multirow}
\usepackage{textcomp}
\usepackage{caption}
\usepackage[ruled,vlined]{algorithm2e}
\usepackage{booktabs,tabularx}
\newcolumntype{C}{>{\centering\arraybackslash}X}

\author{
Zekai Li\textsuperscript{1}\quad
Xiaoyi Fan\textsuperscript{2}\quad
Yifei Zhu\textsuperscript{1}\\[0.5em]
{\normalsize\textsuperscript{1}Shanghai Jiao Tong University, China}\\[0.2em]
{\normalsize\textsuperscript{2}Jiangxing Intelligence, Hong Kong}
}
\hypersetup{pdfauthor={Zekai Li, Xiaoyi Fan, Yifei Zhu},pdftitle={MoS: Mixture-of-Modality Streaming for Low-Latency Online Video LLM Analytics over Dynamic Networks}}
\begin{document}

\date{}

\title{MoS: Mixture-of-Modality Streaming for Low-Latency Online Video LLM Analytics over Dynamic Networks}

\maketitle

\begin{abstract}
The serving model for next-generation video analytics systems is shifting from traditional task-specific vision models to general-purpose online video Large Language Models (LLMs). Online video LLMs enable flexible, contextual understanding of continuously arriving video content via natural language. However, their substantial resource requirements typically mandate cloud-based inference, making continuous video streaming from user devices necessary to keep the cloud model's context current for online interaction. This introduces significant challenges under existing dynamic and often harsh networks, degrading both accuracy and responsiveness in latency-sensitive applications.
In this paper, we propose MoS, a new streaming system that adaptively streams a combination of visual and derived textual modalities over fluctuating networks to enable Pareto-optimal video LLM analytics. The key insight is to leverage the complementarity and interchangeability of different modalities, hidden in video LLM, to improve system performance. Concretely, MoS introduces adaptive anchor frame selection and incremental content tracking to enable low-latency text description generation at the device side. A joint profiler characterizes the accuracy and overhead of paired video and textual configurations, including caption input resolution and output-token budget. A stochastic optimization-based modality controller then selects modality combinations and configurations using this profile and the asynchronous latency of the two modalities to adapt to network conditions.
Across 500 StreamingBench video--question pairs, 4G/5G uplink traces, and the VideoLLM-Online and DisPider backends, MoS reduces latency violations by up to 94.3\% and improves accuracy by up to 60.4\% relative to the evaluated streaming baselines.
\end{abstract}

\section{Introduction}
 Driven by advancements in computer vision and the wide\-spread adoption of networked cameras \cite{xu2023edge}, real-time video analytics, which involves continuous inference over real-time video streams, is rapidly gaining traction.
Traditionally, video analytics services rely on specialized vision models for fixed tasks such as object detection and semantic segmentation~\cite{jiang2018chameleon,zhang2022casva}. Recent general-purpose online video LLMs support diverse natural-language queries over multimodal inputs. While conventional video LLMs analyze a complete, prerecorded video before responding, online models continuously process incoming frames and maintain an evolving context~\cite{Maaz2023VideoChatGPT,chen2024videollm}.
Some online video LLMs can further interact proactively, responding when an observed event warrants user attention rather than waiting for the video to end or for an explicit query~\cite{qian2025dispider}. This capability enables latency-sensitive applications such as wearable visual navigation, smart retail assistance, and robot-guided services, where responses must reflect the latest scene and arrive at the appropriate moment.

\begin{figure}[!t]
    \centering
    \includegraphics[width=\linewidth]{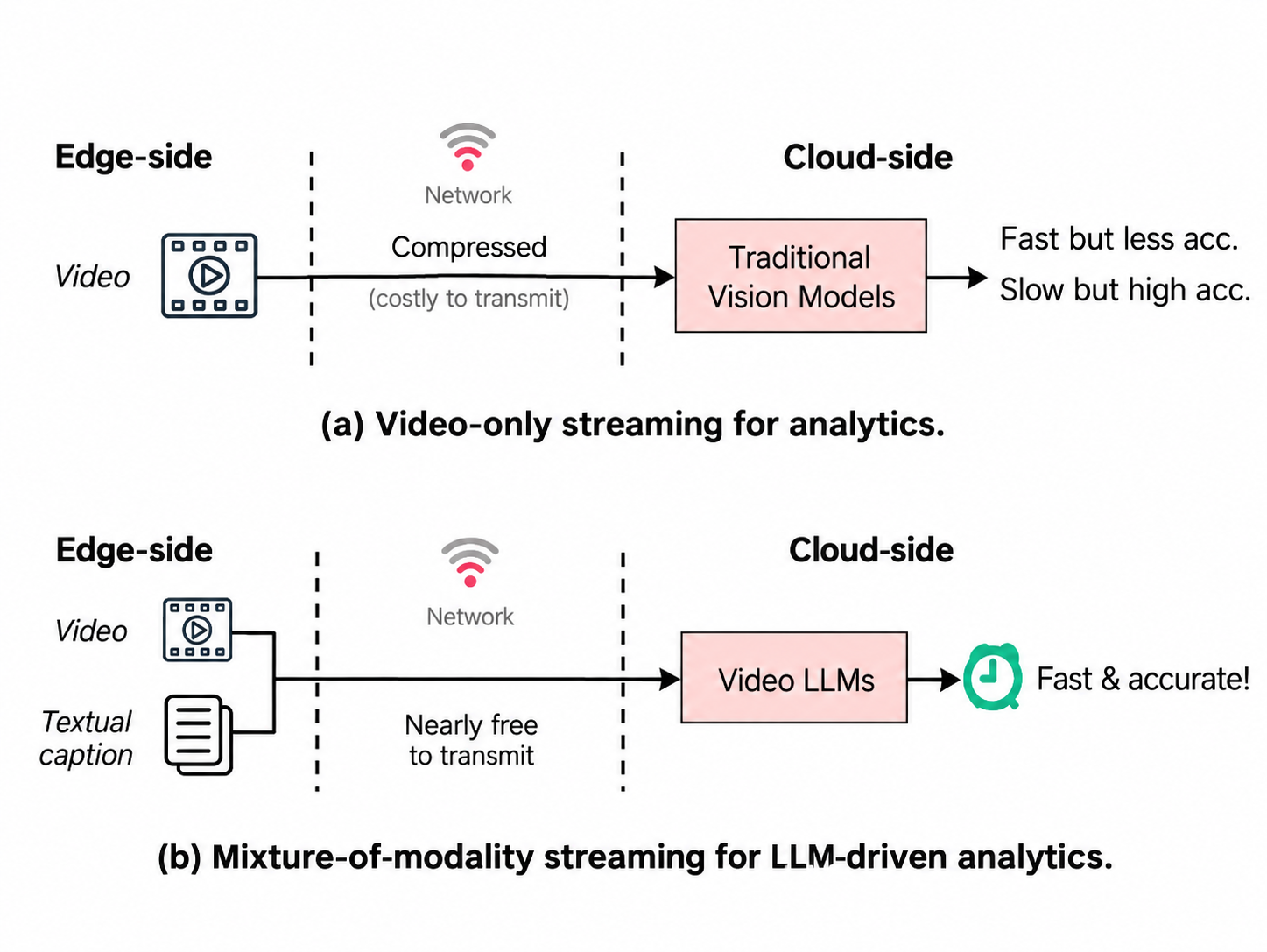}
    \caption{Comparison of two streaming paradigms:
    (a) video-only streaming with forced latency-accuracy tradeoff; (b) mixture of modality streaming with both accuracy and latency ensured.}
    \vspace{-3mm}
    \label{cover}
\end{figure}

MoS targets the recurring operating regime shared by these deployments rather than every video workload: a battery-powered or embedded camera continuously observes a scene, queries arrive online, uplink capacity varies at sub-minute time scales, and a substantially larger cloud model provides the final answer. This regime also appears in remote inspection, body-worn assistance, and field robotics, where transmitting every high-quality frame is expensive but discarding the visual stream entirely is unsafe for fine-grained questions. Our claim is therefore scoped to interactive edge--cloud video LLM services with constrained uplinks; offline batch understanding and fixed-camera deployments with provisioned wired links are outside the intended scope.

Due to the high computational demands of running video LLMs, these services are typically hosted on remote servers or rely on cloud APIs, such as OpenAI’s advanced voice and video platform, to infer results~\cite{openai2024}. To maintain current visual context for online queries and proactive responses, video frames must be transmitted continuously to the cloud for processing, making end-to-end system performance highly sensitive to fluctuations in network bandwidth. As shown in Fig. \ref{cover} (a), when the network is constrained, maintaining high video quality to preserve analytics accuracy often results in significant latency. Conversely, reducing video quality to lower network latency via existing approaches such as adaptive streaming \cite{zhang2022casva}, region-aware encoding \cite{wang2022vabus}, and frame filtering \cite{li2020reducto} inevitably compromises accuracy.

As existing video analytics systems struggle with the tradeoff between accuracy and latency, we observe two key properties of video LLMs that inform better system design: multimodal complementarity and interchangeability. From a system perspective, different modalities complement each other and offer distinct and synergistic benefits for accuracy. For example, the video modality provides rich, fine-grained visual context but incurs high computational overhead. In contrast, the text modality (e.g., compact summaries of objects and events) is faster to process and more focused, yet lacks global scene understanding. When combined, they can answer a broader range of queries more accurately and efficiently. On the other hand, different modalities demonstrate different resilience to network variations and offer partial interchangeability. Video transmission, while informative, is more sensitive to network conditions. When bandwidth is insufficient, textual descriptions can replace video to deliver certain semantic information, although fine-grained visual details may be lost.

This insight motivates us to propose a new streaming paradigm, named Mixture-of-Modality (MoM) streaming. As illustrated in Fig.~\ref{cover}(b), MoM streaming transmits a combination of visual and textual modalities over dynamic networks. It is a model-native design that aims at exploiting the complementary and interchangeable benefits across different modalities for better system design.
While the vision is promising, its systematic realization still poses several non-trivial challenges.
\textit{First, how to generate informative textual captions from video streams without incurring substantial latency on edge devices?}
While textual content is insensitive to network dynamics and incurs minimal transmission overhead, to enable MoM streaming, video frames have to be processed on live to generate a corresponding textual description, introducing extra captioning latency.
Naively captioning every frame could lead to redundant descriptions with unnecessary computation and inference latency. Adopting existing contextual video captioning techniques either relies on resource-intensive models~\cite{zhou2024streaming, xue2025progress} or performs dense captioning on offline videos~\cite{krishna2017dense, yang2023vid2seq}, making them unsuitable for low-latency online streaming scenarios.
\textit{Second, how to stream a combination of video and textual content over dynamic networks for optimized accuracy and latency?}
Fundamentally, the impact of combined modalities on accuracy and overhead in video analytics and the valid control knobs for textual modality streaming all remain underexplored.
Furthermore, jointly adapting two asynchronous modalities on different timelines under a single end-to-end latency SLO makes online modality selection non-trivial.

In this paper, we present MoS, an edge-assisted streaming system that instantiates our proposed mixture-of-modality streaming paradigm to realize efficient real-time video LLM analytics over dynamic networks.
MoS consists of two tightly integrated components. First, a lightweight collaborative textual visual encoder generates rich anchor captions for key frames and incremental captions based on content changes.
Second, a network-aware controller continuously determines which modalities to transmit and at what quality. We first explore valid configuration knobs for both modalities and build a joint profiler capturing their accuracy–overhead tradeoffs. We then formulate modality adaptation as an online constrained optimization problem, solved via a drift-plus-penalty algorithm that explicitly accounts for the asynchronous latency characteristics of the two modalities.
To our knowledge, MoS is the first system to realize mixture-of-modality streaming for real-time video LLM analytics. Our main contributions are as follows:
\begin{itemize}
    \item We conduct a comprehensive analysis of video LLM analytics, revealing the potential to overcome the accuracy-latency tradeoff in traditional video-only streaming via blending different modalities (Sec. \ref{sec:bg}).
    \item We design a collaborative textual visual encoder for fast and informative captions via anchor captioning and motion-aware incremental captioning (Sec. \ref{sec:cap}).
    \item We develop a joint video--text profiler and a stochastic optimization-based controller that adapt modality combinations and video/caption configurations according to their joint accuracy and asynchronous latency, with theoretical long-term latency guarantees (Sec. \ref{sec:control}).
    \item We implement MoS and evaluate 500 StreamingBench pairs over real 4G/5G traces using two backend models. MoS reduces the latency violation rate by up to 94.3\% and improves accuracy by up to 60.4\% relative to the evaluated baselines (Sec.~\ref{sec:eval}).
\end{itemize}

\section{Background \& Motivation}
\label{sec:bg}

\begin{figure}[t]
    \centering
    \begin{minipage}[t]{0.48\linewidth}
        \centering
        \includegraphics[width=\linewidth]{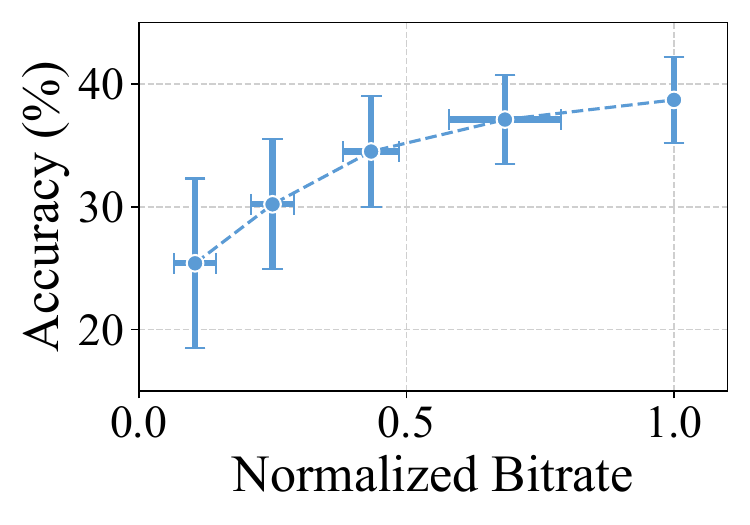}
        \caption{Analytics accuracy versus normalized bitrate at different video quality levels.}
        \label{fig:acc_br}
        \vspace{-5mm}
    \end{minipage}
    \hfill
    \begin{minipage}[t]{0.48\linewidth}
        \centering
        \includegraphics[width=\linewidth]{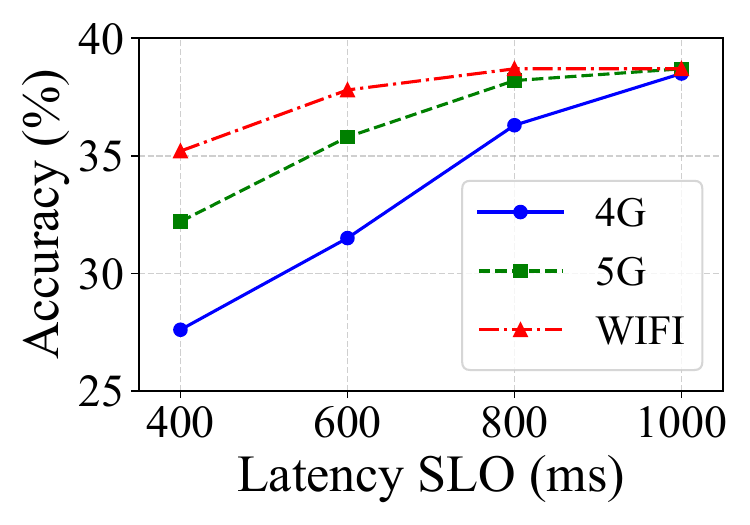}
        \caption{Accuracy under different latency SLOs across varying networks.}
        \label{fig:acc-slo}
        \vspace{-5mm}
    \end{minipage}
\end{figure}

\begin{figure*}[t]
    \centering
    \begin{minipage}[t]{0.31\linewidth}
        \centering
        \includegraphics[width=\linewidth]{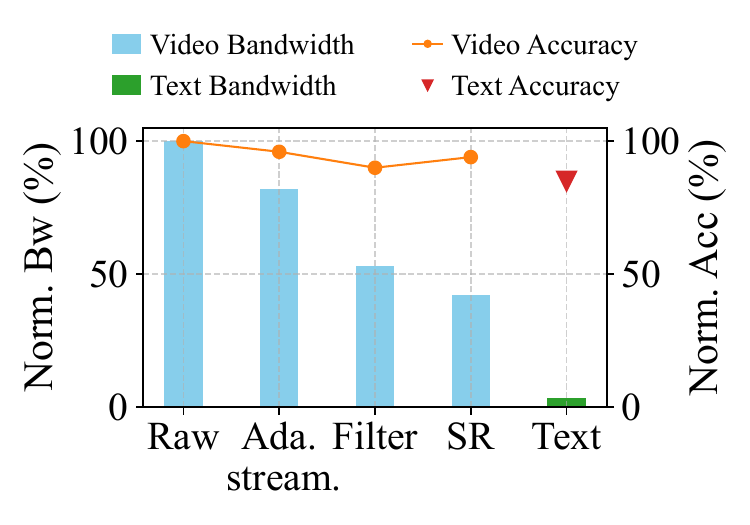}
        \vspace{-5mm}
        \caption{Bandwidth and accuracy of video streaming methods and textual captions.}
        \label{fig:modallim}
        \vspace{-5mm}
    \end{minipage}
    \hfill
    \begin{minipage}[t]{0.3\linewidth}
        \centering
        \includegraphics[width=\linewidth]{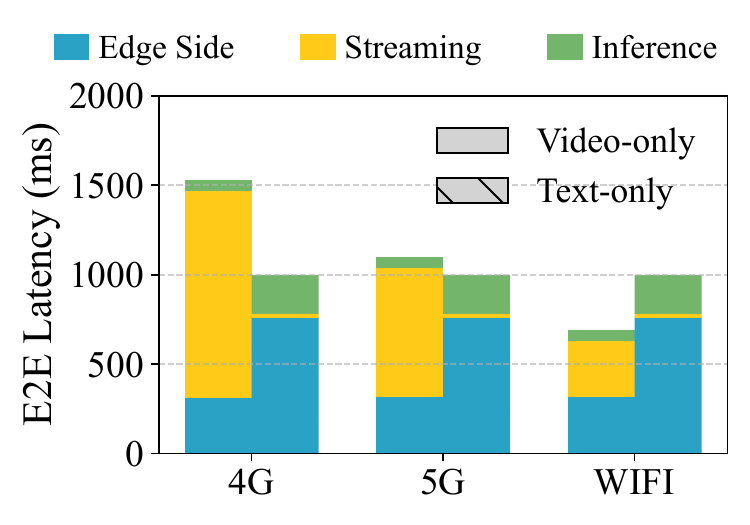}
        \vspace{-5mm}
        \caption{End-to-end latency breakdown for visual and textual modalities.}
        \label{fig:lt-bk}
        \vspace{-5mm}
    \end{minipage}
    \hfill
    \begin{minipage}[t]{0.3\linewidth}
        \centering
        \includegraphics[width=\linewidth]{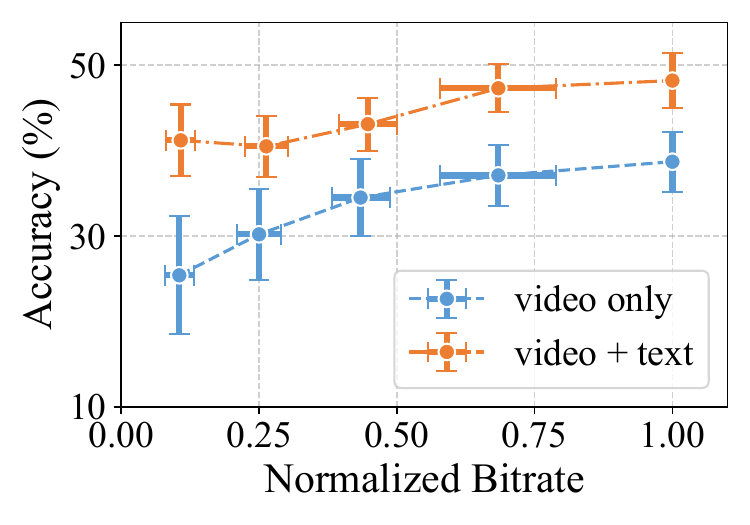}
        \vspace{-5mm}
        \caption{Accuracy and bandwidth of video-only and mixed modalities.}
        \label{fig:acc-ip}
        \vspace{-5mm}
    \end{minipage}
\end{figure*}

\subsection{Online Video Large Language Models}
Traditional video analytics systems~\cite{jiang2018chameleon,zhang2022casva} rely on task-specific vision models for fixed queries such as object detection and semantic segmentation. Video LLMs instead support diverse natural-language queries by jointly reasoning over multimodal inputs. However, conventional video LLMs mostly operate offline: they process a complete, prerecorded video and respond only after the entire input is available.
Recent online video LLMs, such as VideoLLM-Online~\cite{chen2024videollm} and DisPider~\cite{qian2025dispider}, continuously process incoming frames and maintain context as the video unfolds. They support real-time dialogue and can proactively respond when an event warrants interaction, without waiting for the video to end. Consequently, online services must continuously deliver new video content to the model; unlike offline video analysis, they cannot simply upload the complete video in one batch.

Despite their capabilities, these models require substantial resources for deployment and inference. For example, running a full-precision Videollm-online \cite{chen2024videollm}, a typical real-time video LLM, consumes over 30 GB VRAM and hundreds of watts of power. In contrast, the Google Pixel 10 used in our smartphone deployment has a Tensor G5 processor and 12 GB of shared RAM. Consequently, most video LLM-based services are hosted in the cloud, such as OpenAI's Advanced Voice with Video mode~\cite{openai2024}, where video content must be transmitted from local devices to remote servers for analytics.

\subsection{The Tradeoff Wall: Video Streaming in Practical Networks}
\label{tradeoffwall}
Since traditional vision models only process vision input, streaming video remains the default setting. In this context, the dynamic and often harsh network becomes a bottleneck, affecting both analytics timeliness and accuracy.
To better understand this tradeoff, we conduct a measurement study of video LLM analytics via video-only streaming.
Fixed-length segments (e.g., one second) are transmitted from the edge device to a remote server running a real-time video LLM.

\noindent\textbf{Measurements Setup:}
For hardware, the controlled measurement platform uses Jetson Orin Nano, equipped with an NVIDIA Ampere GPU with 1024 CUDA cores. We additionally evaluate a Google Pixel 10 smartphone separately in Sec.~\ref{sec:eval-practicality}. The cloud platform consists of two A800 GPUs and a 26-core Intel Xeon Gold 5320 CPU. We adopt VideoLLM-Online~\cite{chen2024videollm} for streaming video analytics. Data transmission between the device and cloud is based on real-world 4G, 5G uplink traces~\cite{ghoshal2022depth} and WIFI, with average throughputs of 7.6Mbps, 14.7Mbps, and 30.6Mbps, respectively.  For evaluation, we randomly select 50 video question-answer pairs from StreamingBench~\cite{lin2024streamingbench}, covering real-time visual understanding and sequential reasoning tasks. The end-to-end latency SLO is set to 800 ms, including the edge computing, streaming, and inference latency.

We begin by evaluating how video quality affects the accuracy of video LLM-based analytics. Each video is encoded at five quality levels using different combinations of resolution, frame rate, and quantization parameter: [360P, 3FPS, 32], [480P, 5FPS, 30], [540P, 10FPS, 28], [720P, 20FPS, 26], and [1080P, 30FPS, 23]. We measure the mean bitrate and mean accuracy at each level. As shown in Fig.~\ref{fig:acc_br}, analytics accuracy decreases as the normalized bitrate is reduced. For example, a 58.4\% and 75.1\% reduction in bitrate results in a 12.1\% and 26.3\% accuracy drop.
We further evaluate the accuracy under different latency SLOs and network conditions. To meet a specific latency constraint, each segment is encoded at the highest quality that remains within the bandwidth limit. As shown in Fig.~\ref{fig:acc-slo}, reducing the latency target from 1200ms to 600ms leads to an accuracy drop of 10.3\% to 30.2\%, demonstrating the impact of this tradeoff in practice.

\textbf{Insight 1:} \textit{Video-only streaming cannot consistently achieve high accuracy within low-latency requirements under dynamic networks.
This necessitates an alternative streaming paradigm for more efficient video LLM analytics.
}

\subsection{Breaking the Wall: Opportunities from Mixture-of-modality}
\label{sec:oppor}
To assess the feasibility of the proposed MoM streaming, we first evaluate the textual modality, where the edge device converts video content into captions for video LLM analytics. FDVS~\cite{you2024towards} is used to convert video into textual captions for streaming and inference, where VILA-3B~\cite{lin2024vila} serves as the captioning model.
For comparison, we implement several popular techniques, including  CASVA \cite{zhang2022casva} for adaptive streaming, Reducto \cite{li2020reducto} for frame filtering, CloudSeg \cite{wang2019bridging} for super resolution.
As shown in Fig.~\ref{fig:modallim}, reducing bandwidth usage by 18.3\%, 45.7\%, and 55.3\% leads to accuracy drops of 8.2\%, 15.2\%, and 10.1\%, respectively, indicating that video-based methods still face the tradeoff. Although SR achieves better accuracy, it introduces excessive latency as a new tradeoff dimension.
In contrast to these video-based methods, the textual modality reduces the bandwidth usage by $10\times$ with a comparable accuracy.

We next analyze the characteristics of each modality in terms of end-to-end system overhead and analytics performance under different network conditions. The latency is divided into edge-side processing, streaming, and server-side inference. For video, edge-side processing includes encoding, while for text, it involves video captioning using a specific vision-language model.
As shown in Fig. \ref{fig:lt-bk}, video streaming results in highly variable end-to-end latency, ranging from 0.6s to 1.5s depending on network conditions. Streaming accounts for the major delay, contributing 78.2\% under 4G (7.6 Mbps) and 45.3\% under WiFi (30.6 Mbps). In contrast, text streaming maintains a stable latency of around 1.0s across all conditions. Most of the delay (79.3\%) comes from edge-side captioning, with negligible streaming delay. Overall, video streaming incurs lower latency under favorable network conditions but becomes significantly slower than text streaming in other situations.

The apparently counter-intuitive result that text-only can be slower than video-only under WiFi is caused by where the work is performed, not by the number of tokens consumed by the backend. The text path must first run an autoregressive VLM on the edge to produce a caption; this local generation dominates its critical path. The video path uses hardware H.264 encoding and, when uplink bandwidth is high, reaches the server before edge caption generation completes. Backend text inference is indeed cheaper than video inference, but the savings are smaller than the additional edge captioning time in the high-bandwidth case. Under 4G, the ordering reverses because video upload becomes dominant.

\textbf{Insight 2:} \textit{Each modality exhibits distinct computation and transmission patterns. Videos are readily available but costly to transmit. Textual captions are cheap to transmit under all conditions, but require additional edge computation.}

Such distinct computation and transmission patterns of different modalities create opportunities to break the tradeoff boundary of video-only streaming: Captions can serve as a supplement for compressed video under limited bandwidth with minimal transmission overhead. To validate this, we transmit and infer videos at various quality levels, both with and without accompanying captions. As shown in Fig. \ref{fig:acc-ip}, accuracy remains above 0.41 even when the bitrate is reduced by 90.6\%, 61.5\% higher than the video-only setting (0.25) at the same quality.

\textbf{Insight 3:} \textit{Mixed modalities can overcome the accuracy-bandwidth tradeoff of video-only streaming by improving analytics accuracy without additional transmission cost.}

\section{MoS Design}

Motivated by the potential of mixture-of-modality streaming, we propose MoS, an edge–assisted streaming system for efficient real-time video LLM analytics. This section outlines the design goals and challenges of MoS (Sec. \ref{sec:goal}), provides a system overview (Sec. \ref{sec:over}), and details the design of each component (Sec. \ref{sec:cap}, \ref{sec:control}).

\subsection{Design Goals and Challenges}
\label{sec:goal}

\noindent\textbf{Goals:}
MoS is designed to address two key objectives in real-time video analytics: \textbf{(i)} maximizing analytics accuracy and \textbf{(ii)} ensuring end-to-end latency meets the target SLO under dynamic network conditions. However, designing MoS involves two key challenges that must be addressed.

\noindent\textbf{Challenge 1:} \textit{How to generate informative textual captions from real-time video streams without significant latency on resource-constrained edge devices? }
Although textual captions offer an alternative to video data with minimal bandwidth usage, generating them with current resource-intensive and mainly offline methods \cite{zhou2024streaming, xue2025progress} could introduce considerable overhead (e.g., 79.3\% of E2E latency).
In addition, applying dense captioning to every streamed frame produces redundant descriptions, which challenges its practicality.

\noindent\textbf{Challenge 2:} \textit{How can video and textual modality streaming be adaptively balanced under highly dynamic network conditions?}
While video-only adaptive streaming has been well studied, the impact of mixed modality configurations remains underexplored, which is essential for effective streaming control. Meanwhile, it is challenging to ensure the controller remains adaptable when video content or network conditions shift, as video LLMs are applied in diverse scenarios.

\begin{figure}[!t]
    \centering
    \includegraphics[width=\linewidth]{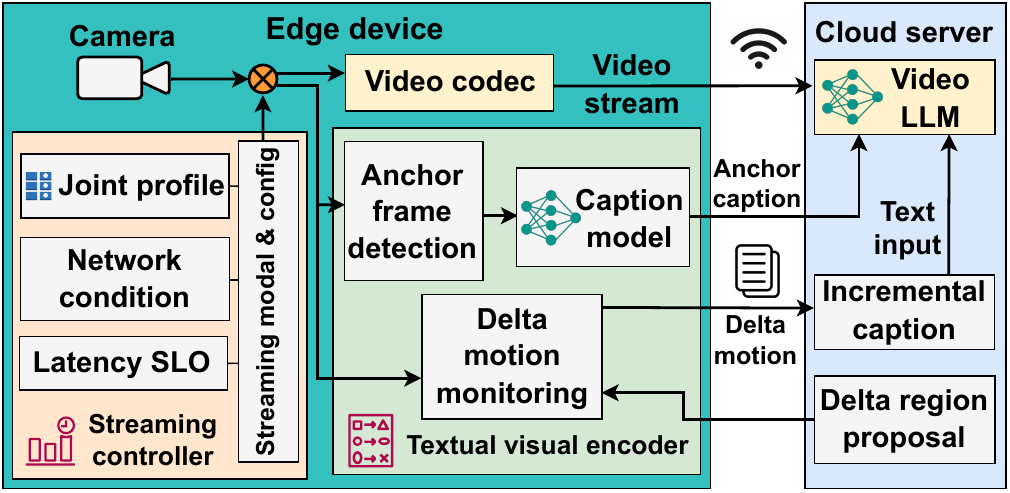}
    \vspace{-4mm}
    \caption{Design overview of MoS.}
    \vspace{-3mm}
    \label{system}
\end{figure}

\subsection{MoS Overview}
\label{sec:over}
To address the outlined challenges, MoS integrates two core components:
\textit{(i)} \textbf{Collaborative Textual Encoder for Video Streaming}, which enables efficient streamed video captioning on edge devices.
To reduce edge-side computation, it selects anchor frames based on accumulated content differences and applies a lightweight vision-language model to caption only these frames, rather than performing dense captioning across the entire video. To capture cross-frame dynamics, motion vectors are extracted and sent to the server. A server-side language model then generates incremental captions using the anchor captions and motion vectors.
\textit{(ii)} \textbf{Adaptive Multimodal Streaming Controller}, which adapts streaming modality and configurations in response to network conditions. It jointly profiles accuracy and system overhead across video-text configuration pairs, then develops an online drift-plus-penalty algorithm to select the optimal streaming mode and quality that maximizes accuracy within latency constraints.

As illustrated in Fig. \ref{system}, the workflow proceeds as follows:
for each segment, the controller selects the modality and configuration for streaming based on video-text joint profile, latency SLO, and network condition. It selects from three candidate modalities: \textit{video-only}, \textit{text-only}, and \textit{video-text}.
If text modality is selected, the textual visual encoder generates anchor captions and records motion information. The encoded data is transmitted over the network to the server, where incremental captions are finally generated. The video LLM then performs continuous inference on the received inputs.

\subsection{Collaborative Textual Visual Encoder}
\label{sec:cap}

\begin{figure}[!t]
    \centering
    \includegraphics[width=\linewidth]{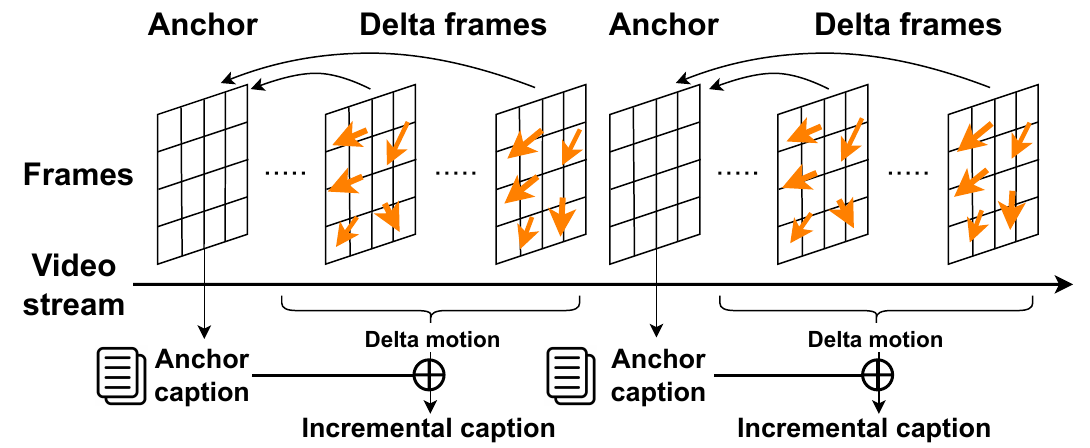}
    \caption{Workflow of anchor-incremental captioning.}
    \vspace{-4mm}
    \label{caption}
\end{figure}

To address \textit{Challenge 1}, we design a collaborative textual visual encoder, which operates in three main steps for each segment as illustrated in Fig. \ref{caption}:
On the edge side, the system monitors frame-wise content differences.
\textit{(i)} If the accumulated difference exceeds a profiled threshold, a vision-language model is triggered to generate an anchor caption for the current frame.
\textit{(ii)} For frames between anchors (delta frames), it records and transmits motion vectors within the specific regions.
\textit{(iii)} A server-side fast language model then generates incremental captions for the delta frames using the latest anchor caption and the motion vectors, capturing cross-frame content changes.

\subsubsection{Anchor Detection \& Captioning}
Running dense video captioning models on all captured frames or the whole video segment at the edge is impractical.
However, occasional captioning of individual frames with a lightweight vision-language model is feasible.
For instance, Jetson Nano can run lightweight models such as the VILA series~\cite{lin2024vila} at approximately 2.7 FPS using only 3 GB of memory~\cite{jetson2024tutorial}. Moreover, adjacent frames in streaming video are often highly similar. We observe that captioning consecutive frames typically yields nearly identical outputs, indicating that dense captioning is both redundant and inefficient.

Building on this observation, we propose selecting only anchor frames with significant content changes for captioning. Prior methods typically use uniform sampling or tools like PySceneDetect \cite{scenedetect} to extract key frames of arbitrary length. However, uniform sampling may miss important content, especially in long videos, while other methods require decoding all frames, making it inefficient for real-time applications.
To address these limitations, we design an efficient anchor frame selection method based on accumulated content differences.
Specifically, we compute the pixel-level difference $D_t$ between adjacent frames $F_t, F_{t+1}$ using mean squared error $MSE(F_t, F_{t+1})$.
For continuous frames, the accumulated difference up to $F_t$ is calculated using an exponentially weighted moving average (EWMA), enabling robust detection of step-like scene changes through a fast single-pass process:
\begin{equation}
    S_t = \alpha D_t + (1-\alpha)S_{t-1},
\end{equation}
where $0< \alpha \le 1$ is a smoothing constant (e.g., 0.4 by default). This formulation depends only on the previous value $S_{t-1}$, resulting in $O(1)$ time and memory complexity, which is well suited for real-time video processing.

\begin{figure}[t]
    \centering
    \begin{minipage}[t]{0.48\linewidth}
        \centering
        \includegraphics[width=\linewidth]{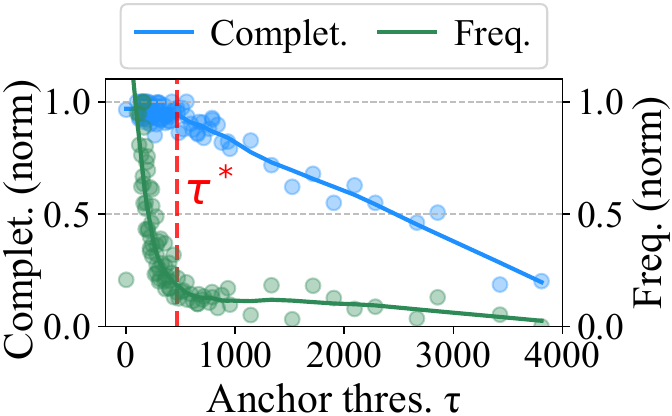}
        \caption{Anchor threshold tuning based on caption completeness and frequency.}
        \vspace{-3mm}
        \label{tune}
    \end{minipage}
    \hfill
    \begin{minipage}[t]{0.48\linewidth}
        \centering
        \includegraphics[width=\linewidth]{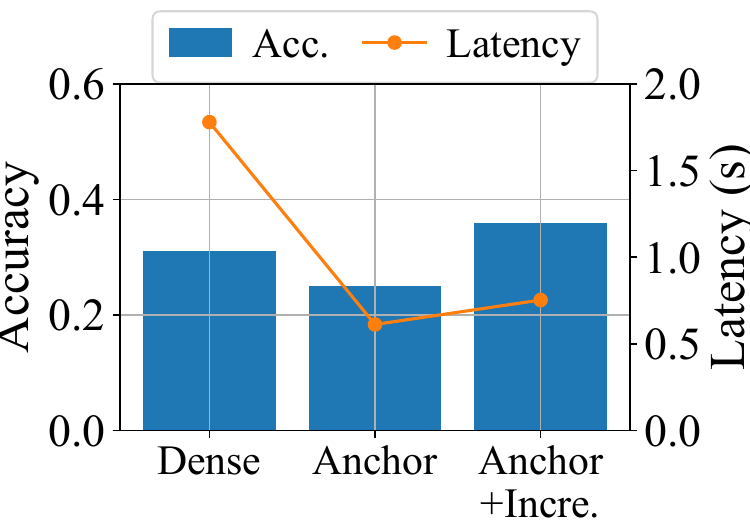}
        \caption{Accuracy and overhead comparison of different caption designs.}
        \vspace{-3mm}
        \label{cp_comp}
    \end{minipage}
\end{figure}

Once the accumulated difference \( S_t \) exceeds a threshold \( \tau \), the corresponding frame \( F_t \) is selected as an anchor frame. This triggers a lightweight vision-language model on the edge device to generate a caption and resets \( S_t \).
However, selecting an appropriate value for \( \tau \) is challenging.
A small \( \tau \) results in frequent and redundant captioning, increasing computational overhead, while a large \( \tau \) risks missing important content. To balance these factors, we design an optimization method to identify a proper threshold that produces informative anchor captions with minimal captioning overhead.
Specifically, for each $\tau$, we evaluate the semantic completeness and frequency of captions across a set of video samples.
To evaluate the completeness of sparse anchor captions \(\mathcal C^{\tau}=\{c^{\tau}_1,\dots,c^{\tau}_{N(\tau)}\}\) given threshold \(\tau\), we compare them against dense reference captions \(\mathcal C^{\text{ref}}=\{c^{\text{ref}}_1,\dots,c^{\text{ref}}_{T}\}\), obtained by applying the same captioner to every frame. Each caption is mapped to a semantic embedding using a sentence encoder \(\mathbf e(\cdot)\). For each reference caption \(c^{\text{ref}}_i\), we compute its maximum cosine similarity across all anchor captions, and define the semantic completeness as:
\begin{equation}
  \operatorname{comp}(\tau)=
  \frac{1}{T}\sum_{i=1}^{T}
  \max_{c \in \mathcal C^{\tau}}
  \operatorname{cos}\!\bigl(
     \mathbf e(c^{\text{ref}}_i),\,
     \mathbf e(c)
  \bigr),
  \label{eq:completeness}
\end{equation}
where a higher $\operatorname{comp}(\tau) \in [0,1]$ indicates a higher semantic completeness. Meanwhile, to penalize redundant anchor captions, we measure the captioning frequency $freq(\tau)$ as the average number of captioning times per second, and fold it into the optimization objective as:
\begin{equation}
    \tau^{*} = \arg\max_{\tau} \, comp(\tau) - freq(\tau).
\end{equation}
As shown in Fig.~\ref{tune}, we report semantic completeness and captioning frequency across different $\tau$. The optimized $\tau$ achieves over 0.9 completeness with a frequency of around 0.47 FPS. Notably, both metrics can be computed without ground-truth labels, enabling continuous adaptation at runtime.

\subsubsection{Motion-aware Incremental Captioning}
While anchor captions informatively describe static visual scenes, they omit cross-frame motions and events. To address this, we propose an additional motion-aware incremental caption method via edge–cloud collaboration.
It extracts motion until the next anchor frame and transmits it textually to the server, where a fast language model generates incremental captions based on the motion and the latest anchor caption.
However, simply computing motion vectors across the entire frame is ineffective, as raw motion data lacks semantic meaning and often includes irrelevant information. It fails to convey what moved, in which direction, and with what significance.

To address this, we first identify objects of interest (OoIs) on the server by leveraging semantic cues from anchor captions. Given a frame \( F_t \) and the latest anchor caption \( C_t \), we extract a list \( L \) of object keywords (e.g., `man', `laptop') from \( C_t \) using a sentence parser. Then for each object \( l_i \in L \), we detect its location in \( F_t \) as a bounding box \( bbox_i \) through visual grounding (e.g., GroundingDINO~\cite{liu2024grounding}). The results are formatted as a list of pairs \( (l_i, bbox_i) \) and shared with the edge device.
This information can be continuously updated as new frames or anchor captions become available. Since the extraction runs on the server and completes within 50ms in our setup, it ensures timely delivery of OoI data during runtime.

\begin{table}[t]
  \centering
  \footnotesize
  \setlength{\tabcolsep}{3pt}
  \renewcommand{\arraystretch}{0.95}
  \caption{Prompt templates used for captioning}
  \label{tab:prompts}
  \begin{tabularx}{\linewidth}{@{}>{\raggedright\arraybackslash}p{0.16\linewidth}|>{\raggedright\arraybackslash}X@{}}
    \toprule
    \textbf{Type} & \textbf{Template} \\ \midrule
    Anchor Caption &
    Describe this frame within \texttt{\textless MAXTOK\textgreater} tokens, covering salient objects, notable attributes, and scene context. \\ \midrule
    Incremental Caption &
    Using the anchor caption \texttt{\textless ANCHOR\textgreater} and the existing narrative \texttt{\textless CONTEXT\textgreater}, read the current object movements \texttt{\textless DELTA\textgreater} and state what has occurred since the last update without repeating previous details. \\ \bottomrule
  \end{tabularx}
  \vspace{-3mm}
\end{table}

After semantic localization, we monitor the motion of OoIs across frames on the edge.
Rather than employing heavy tracking models, we leverage motion vectors from the video codec during video encoding, which requires no extra computation.
Modern encoders expose \(16\times16\) macro-block motion vectors during encoding. For each object \( l_i \), we collect the motion vectors within its bounding box \( bbox_i \), compute the median values \( \Delta x, \Delta y \) as the motion direction, and shift the bounding box accordingly. The motion \( M_t \) between frames \( F_t \) and \( F_{t+1} \) is then represented as \( (l_i, bbox_i, \Delta x_i, \Delta y_i) \).
When a new anchor frame is detected or a segment is completed, the recorded motion is formatted into a list of \( M_t \) and sent to the server. A fast language model generates incremental captions based on the latest anchor caption, the motion, and any previous incremental captions. This server-based process adds minimal latency, typically around 100ms. The detailed prompt template for caption is shown in Table \ref{tab:prompts}.

\subsection{Multimodal Streaming Controller}
\label{sec:control}

To address \textit{Challenge 2}, we introduce an \textbf{Adaptive Multimodal Streaming Controller} that jointly decides \emph{which modality to stream} and \emph{at what quality}. This decision is fundamentally different from conventional video adaptation. Video and textual captions exhibit distinct computation--communication tradeoffs: increasing video quality mainly increases encoding and transmission costs, whereas increasing caption quality changes both local captioning latency and textual content transmitted to the server. Moreover, when the two modalities are used together, their pipelines proceed largely in parallel and backend inference can start only after both inputs become available. Therefore, independently adapting the two modalities or selecting them using a single bitrate-like metric can easily lead to suboptimal decisions.

Our controller addresses this problem in two steps. First, a \emph{mixture-of-modality joint profiler} characterizes the task accuracy and system overhead of different modality-quality combinations. Second, at runtime, a network-aware controller combines the profile with current bandwidth conditions and recent latency violations to select a configuration for every segment. Rather than relying solely on instantaneous bandwidth prediction, the controller maintains a virtual latency queue that records accumulated SLO ``debt.'' This allows it to react not only to the predicted latency of the current segment, but also to persistent prediction errors and recent SLO violations.

\subsubsection{Mixture-of-Modality Joint Profiler}

Selecting a streaming configuration requires jointly reasoning about three factors: task accuracy, computation overhead, and communication cost. The difficulty is that these factors vary differently across modalities. A higher-quality video increases the amount of visual evidence available to the backend model but also increases encoding and upload costs. In contrast, textual captions require local model execution before transmission, and their quality depends on both the visual input provided to the captioner and the amount of text it is allowed to generate. Consequently, there is no single quality knob that can directly compare video-only, caption-only, and multimodal streaming configurations.

\noindent\textbf{Joint configuration space.}
For video streaming, we consider $(N+1)$ candidates
$\{V_0,\ldots,V_N\}$, where each $V_n$ is a tuple of video resolution,
frame rate, and quantization parameter. We additionally define $V_0$ as
the zero-quality configuration, indicating that no video is transmitted.

Compared with video, the configuration space for textual captions is less
well established in streaming systems. Under a fixed captioning prompt,
both captioning latency and caption detail are primarily determined by two
factors: the amount of visual information processed by the caption model
and the maximum number of output tokens. We therefore expose two knobs:
\emph{input frame resolution} and \emph{maximum output token length}.
The former controls the amount of visual input processed by the captioner,
while the latter directly bounds generation cost and caption length. These
knobs are also directly supported by commonly used multimodal inference
frameworks such as Ollama and vLLM.

Accordingly, we construct $(M+1)$ caption configurations
$\{T_0,\ldots,T_M\}$, where each $T_m$ specifies an input frame resolution
and maximum output token length, and $T_0$ disables caption generation.
The complete action space is then the Cartesian product
\begin{equation}
\mathcal J = \{J_{n,m}=(V_n,T_m)\mid
0\le n\le N,\;0\le m\le M\}.
\end{equation}
This representation unifies modality selection and quality adaptation in a
single decision space. Specifically, $(V_0,T_m)$ corresponds to
caption-only streaming, $(V_n,T_0)$ to video-only streaming, and
$(V_n,T_m)$ with $n,m>0$ to joint multimodal streaming. Thus, switching
between modalities does not require a separate decision rule; it naturally
emerges from configuration selection.

\noindent\textbf{Offline bootstrap profiling.}
The relationship between a joint configuration and downstream performance
is highly non-linear~\cite{zhang2018awstream}. For example, reducing video
resolution can substantially decrease transmission latency while having
little effect on some tasks, whereas on fine-grained recognition tasks the
same change may cause a large accuracy loss. Similarly, a longer caption
does not necessarily provide proportionally more useful information.
Instead of analytically modeling these effects, MoS adopts a data-driven
profiler using representative video samples such as those described in
Sec.~\ref{tradeoffwall}.

For every configuration $J\in\mathcal J$, the profiler records
\begin{equation}
\mathcal P(J)=
\left\{
Acc(J),\,Bw_v(J),\,Bw_c(J),\,
T_{\mathrm{cap}}(J),\,T_{\mathrm{enc}}(J),\,T_{\mathrm{inf}}(J)
\right\},
\label{eq:profile}
\end{equation}
where $Acc(J)$ denotes the downstream analytics accuracy;
$Bw_v(J)$ and $Bw_c(J)$ are the video and caption payload sizes per
segment; and $T_{\mathrm{cap}}(J)$, $T_{\mathrm{enc}}(J)$, and
$T_{\mathrm{inf}}(J)$ denote caption generation, video encoding, and
backend inference latency, respectively. Importantly, the profiler uses
\emph{task accuracy} rather than a modality-specific proxy such as PSNR or
caption similarity, allowing configurations from different modalities to
be compared using the same objective.

\noindent\textbf{Online profile refinement.}
A purely offline profile can become stale when the scene distribution,
device load, or workload changes. MoS therefore treats the offline table as
a bootstrap rather than a static oracle and continuously calibrates it
during execution.
Raw, uncompressed video is used as the reference for online profiling.
When raw video is already transmitted, the server can directly reuse it to
evaluate alternative configurations. Otherwise, MoS opportunistically
uses spare network capacity to upload occasional raw segments for
re-profiling. Profiling is performed asynchronously on the server and can
exploit configuration-level parallelism and temporal sampling techniques
well explored in prior adaptive video systems~\cite{zhang2018awstream}.
As a result, profile refresh does not lie on the critical streaming path.
This offline--online co-design keeps the profiling error within 5\% over
time while avoiding continuous raw-video transmission.

\subsubsection{Network-Aware Configuration Adaptation}

The joint profile describes the intrinsic cost and accuracy of each
configuration, but it does not determine which configuration should be
used at runtime. The same configuration can easily meet a latency target
under a high-bandwidth connection and violate it when the network
degrades. Meanwhile, bandwidth prediction itself is imperfect.
Selecting the highest-accuracy configuration that merely \emph{appears}
feasible can therefore repeatedly accumulate SLO
violations.

We address this with a closed-loop controller combining two complementary
mechanisms. A \emph{predictive feasibility gate} first avoids
configurations that are already expected to violate the current SLO.
Among the remaining choices, a \emph{virtual latency queue} tracks
previous violations and gradually makes the controller more conservative
when latency debt accumulates. Finally, the queue is updated using
\emph{measured}, rather than predicted, end-to-end latency. This feedback
is important in practice: prediction determines the current action, while
observation corrects future actions when the prediction is inaccurate.

\noindent\textbf{Causal bandwidth prediction.}
The prototype uses a one-segment-ahead predictor with the same horizon as the one-second control interval. At the start of segment $t$, it computes $\hat B_t$ as the harmonic mean of the goodput observed for the five most recently \emph{completed} uploads, after clipping each sample to the 5th--95th percentile range of the same causal window. No throughput sample from segment $t$ or the future is available to the decision. A harmonic mean is intentionally conservative for upload-time estimation and is less dominated by brief high-rate bursts than an arithmetic mean. Appendix~\ref{app:bw-error} reports its trace-replay error, and Sec.~\ref{sec:eval-robustness} evaluates the controller on the original correlated mobility traces; the analytical i.i.d. assumption is not used to preprocess or shuffle those traces.

\noindent\textbf{Latency estimation.}
For segment $t$, let $\hat B_t$ denote the estimated available bandwidth.
Caption generation and video encoding run independently, after which their
outputs are transmitted to the server. Since backend inference requires
the selected modalities to be available, the pre-inference critical path
is determined by the slower branch. For configuration $J_t$, we estimate
its end-to-end latency as
\begin{equation}
\begin{aligned}
\hat T_{\mathrm{tot}}(J_t,\hat B_t)
={}&\max\Bigl\{
T_{\mathrm{cap}}(J_t)+\frac{Bw_c(J_t)}{\hat B_t},\\
&\qquad T_{\mathrm{enc}}(J_t)+\frac{Bw_v(J_t)}{\hat B_t}
\Bigr\}\\
&+T_{\mathrm{inf}}(J_t).
\end{aligned}
\label{eq:lat-est}
\end{equation}
For video-only or caption-only configurations, the disabled branch has
zero computation and transmission cost and therefore naturally disappears
from Eq.~\eqref{eq:lat-est}. In practice, caption transmission is usually
much smaller than caption generation and video upload, but we retain it to
keep the estimator valid across different caption lengths and network
conditions.

\noindent\textbf{Problem formulation.}
Our objective is to select a joint configuration $J_t\in\mathcal J$ for
each segment to maximize long-term task accuracy while satisfying an
average end-to-end latency target $T_{\mathrm{SLO}}$:
\begin{align}
\max_{\{J_t\}}
\quad&
\liminf_{N\rightarrow\infty}
\frac{1}{N}
\sum_{t=1}^{N}
\mathbb E\!\left[\mathrm{Acc}(J_t)\right]
\nonumber\\[3pt]
\text{s.t.}\quad&
\limsup_{N\rightarrow\infty}
\frac{1}{N}
\sum_{t=1}^{N}
\mathbb E\!\left[
T_{\mathrm{tot}}(J_t,B_t)-T_{\mathrm{SLO}}
\right]
\le 0,
\label{eq:cmdp}
\end{align}
where $B_t$ denotes the realized bandwidth and
$T_{\mathrm{tot}}(J_t,B_t)$ is the corresponding observed end-to-end
latency.

We deliberately formulate the constraint over long-term latency rather
than requiring every individual segment to satisfy the SLO. A strict
per-segment constraint is brittle under short bandwidth drops and
prediction errors: even the lowest-cost configuration may occasionally
miss the target. Instead, MoS combines a per-segment \emph{predicted}
feasibility check with long-term feedback from realized latency. The former
prevents obviously unsafe decisions, while the latter prevents repeated
small prediction errors from accumulating unnoticed.

\noindent\textbf{Virtual latency queue.}
To capture historical SLO violations, we maintain a virtual queue
$Q_t\ge0$. After executing segment $t$, the controller observes the actual
end-to-end latency $T_{\mathrm{obs},t}$ and updates
\begin{equation}
Q_{t+1}
=
\max\!\left\{
0,\;
Q_t+T_{\mathrm{obs},t}-T_{\mathrm{SLO}}
\right\}.
\label{eq:queue-update}
\end{equation}

The queue has a simple interpretation as accumulated \emph{latency debt}.
A segment faster than the SLO reduces the debt, whereas an SLO violation
increases it. Therefore, two segments experiencing the same current
bandwidth can result in different configuration choices: after a sequence
of successful low-latency segments, the controller can afford a more
aggressive high-accuracy configuration; after recent violations, the same
configuration receives a larger latency penalty.

This history dependence is particularly useful when bandwidth prediction
contains systematic or transient errors. Instead of explicitly modeling
prediction uncertainty, the observed latency is fed back through $Q_t$,
providing a lightweight integral correction mechanism.

\noindent\textbf{Drift-plus-penalty selection.}
At the beginning of segment $t$, the controller first constructs the
predicted-feasible set
\begin{equation}
\mathcal J_t^{\mathrm{safe}}
=
\left\{
j\in\mathcal J:
\hat T_{\mathrm{tot}}(j,\hat B_t)
\le T_{\mathrm{SLO}}
\right\}.
\end{equation}
This first-stage filtering removes configurations that would knowingly
violate the SLO under the current network estimate.

Among the remaining candidates, MoS applies a drift-plus-penalty rule and
selects
\begin{equation}
J_t^\star
=
\arg\max_{j\in\mathcal J_t^{\mathrm{safe}}}
\left[
V\,\mathrm{Acc}(j)
-
Q_t\,\hat T_{\mathrm{tot}}(j,\hat B_t)
\right],
\label{eq:dpp-obj}
\end{equation}
where $V>0$ controls the accuracy--latency tradeoff. The two terms have
distinct roles. $V\,\mathrm{Acc}(j)$ rewards configurations with higher
analytics quality, whereas
$Q_t\,\hat T_{\mathrm{tot}}(j,\hat B_t)$ makes expensive configurations
increasingly unattractive as latency debt grows. A larger $V$ therefore
makes the controller more accuracy-oriented, while a larger $Q_t$
automatically shifts preference toward configurations with shorter
critical paths.

This design differs from simply adding a fixed latency penalty to
accuracy. A fixed penalty requires choosing one operating point in
advance and cannot adapt its aggressiveness after runtime violations. In
contrast, $Q_t$ is state dependent: the latency penalty is weak when the
system has recently operated safely and becomes progressively stronger
when SLO debt accumulates.

If $\mathcal J_t^{\mathrm{safe}}$ is empty, the SLO is predicted to be
temporarily infeasible. Rather than stalling or selecting an arbitrary
configuration, MoS enters a recovery mode and chooses
\begin{equation}
J_t^\star
=
\arg\min_{j\in\mathcal J}
\hat T_{\mathrm{tot}}(j,\hat B_t).
\end{equation}
This fallback minimizes the expected additional latency debt and drives the
system back toward the feasible region as quickly as possible.

Alg.~\ref{alg:cfg-ctrl} in Appendix~\ref{app:algorithm} summarizes the complete procedure.

\begin{figure}[!t]
\centering
\includegraphics[width=\linewidth]{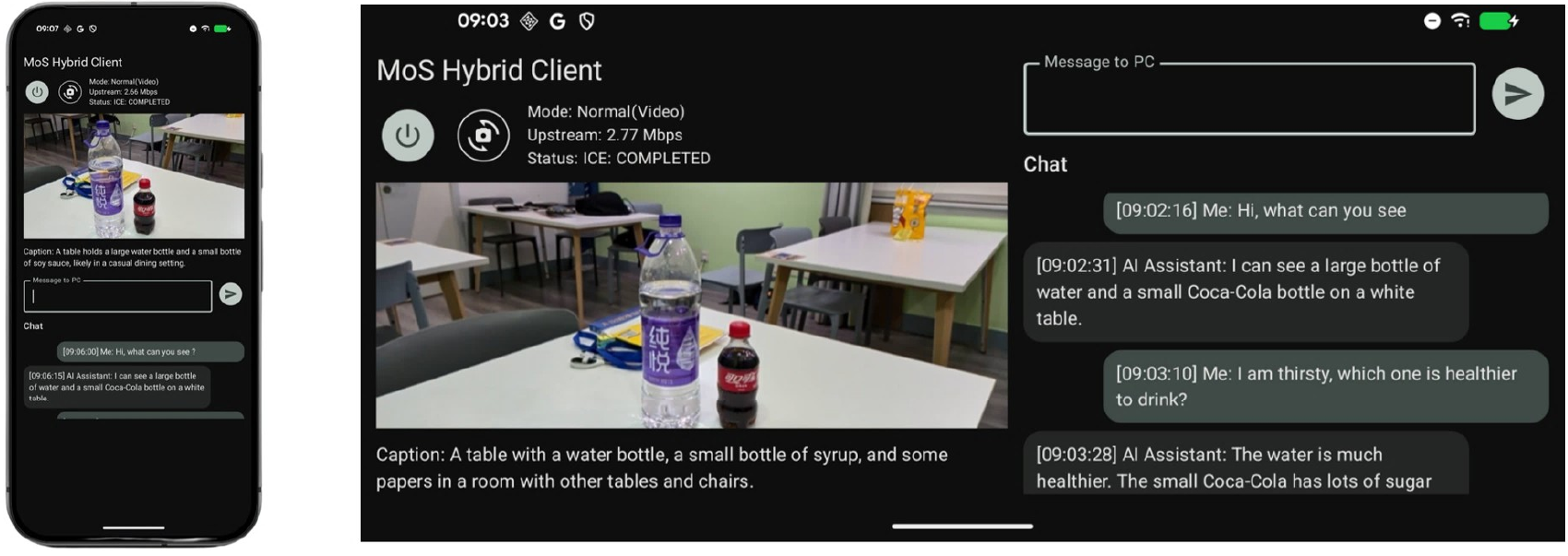}
\caption{Device and interface of MoS's real-world prototype.}
\vspace{-3mm}
\label{demo}
\end{figure}

\begin{figure*}[!t]
\centering
{\includegraphics[width=\textwidth]{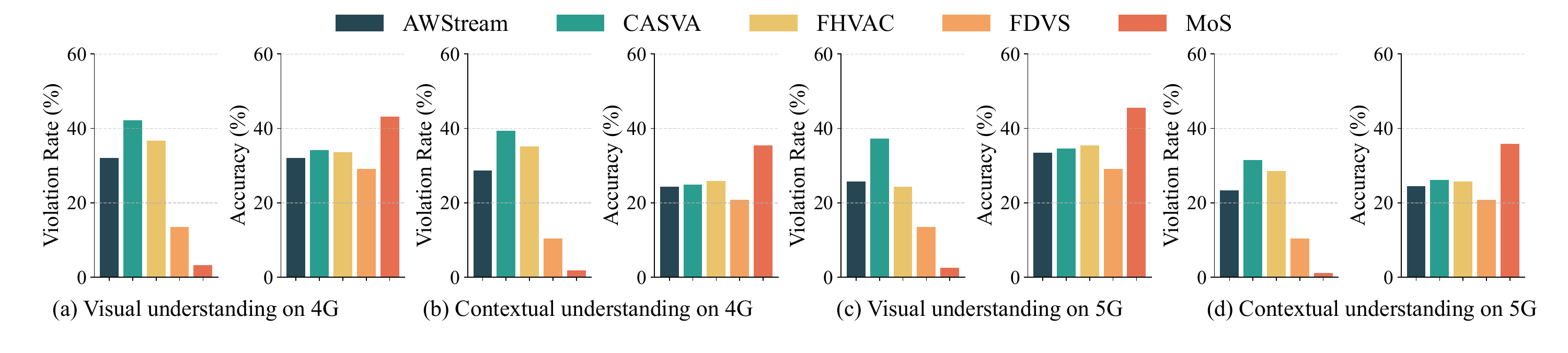}}
\vspace{-6mm}
\caption{The overall performance of VideoLLM-online under different network conditions and tasks.}
\vspace{-4mm}
\label{fig:overall}
\end{figure*}

The performance guarantee, theorem, proof, and applicability conditions are provided in Appendix~\ref{app:guarantee}.

\subsection{Implementation and Deployment}
The MoS implementation comprises approximately 3.7K lines of Python code (LoC), distributed as follows: 18.7\% for real-time video encoding and streaming, 23.6\% for the textual visual encoder, 31.9\% for the adaptive multimodal streaming controller's profiling and online control, 21.2\% for real-time video LLM inference interfaces, and 4.6\% for supporting utilities and infrastructure.

The Jetson testbed uses the platforms described in Sec.~\ref{tradeoffwall}, while Fig.~\ref{demo} shows the smartphone client. To evaluate deployment on a commodity smartphone, we additionally port the edge client to a Google Pixel 10 equipped with a Tensor G5 processor and 12~GB of RAM. The server runs Ubuntu 20.04 LTS with PyTorch 2.2.0. Real-time streaming from the edge device to the server uses UDP sockets. Video processing and encoding use OpenCV~\cite{opencv_python} and FFmpeg~\cite{ffmpeg_python} with the H.264/MPEG-4 codec. The video LLM interface is built on VideoLLM-Online-8B-v1+~\cite{chen2024videollm}. We use VILA-1.5-3B-4bit~\cite{lin2024vila} as the edge-side vision-language model and LLaMA3.1~\cite{touvron2023llama} for incremental caption generation.

On Jetson, video encoding uses the hardware NVENC path and captioning runs as a persistent quantized GPU process so weights are loaded once rather than per segment. On Pixel, camera capture and H.264 encoding use Android's hardware media codec; the caption model is packaged as a 4-bit mobile runtime asset and invoked through the device accelerator when supported, with GPU/CPU fallback. The streaming client exposes the same RPC schema on both devices, so the server-side grounding, incremental-captioning, and video-LLM services are unchanged. Each request carries a stream identifier, segment timestamp, configuration identifier, and byte counts for the video and text branches. These fields let the server reject stale segments, align asynchronous modalities, and account for anchor thumbnails and motion records. Model files are cached locally, payloads can be encrypted by the transport layer, and raw frames are discarded after acknowledgement unless an explicit profiling sample is retained. Appendix~\ref{app:devices} reports sustained power, memory, and thermal behavior instead of relying only on single-query latency.

\section{Evaluation}
\label{sec:eval}

We evaluate MoS by answering the following questions:
\begin{enumerate}
    \item How effectively does MoS improve analytics accuracy while meeting latency SLOs across tasks and backend model stacks? (\S\ref{sec:eval-overall})
    \item Can MoS adapt to varying latency SLOs and network conditions and generalize to unseen deployments and tasks? (\S\ref{sec:eval-robustness})
    \item How do modality selection and the core system components contribute to MoS's performance? (\S\ref{sec:eval-design})
    \item Is MoS practical under edge resource constraints, and what runtime overhead does it introduce? (\S\ref{sec:eval-practicality})
\end{enumerate}

The key takeaways are: (1) \textbf{Higher accuracy with fewer SLO violations:} Across two backend model stacks, MoS improves average accuracy by 32.0--60.4\% over the baselines while reducing violation rates by 81.8--94.3\%. (2) \textbf{Robust adaptation and generalization:} MoS maintains the best accuracy--violation tradeoff across latency SLOs from 400 to 1000\,ms and mean bandwidths from 2 to 40\,Mbps, and retains its advantage on a real-world 5G trace and an unseen analytics task. (3) \textbf{Effective multimodal coordination:} The online controller is critical to avoiding SLO violations under runtime shifts, while the mixed video--text mode substantially improves accuracy; adding an embedding stream provides no further accuracy benefit. (4) \textbf{Practical edge--cloud execution:} MoS keeps average end-to-end latency below 800\,ms with about 3\,GB of edge VRAM and below 40\% CPU utilization, whereas quantized edge-only inference violates latency or memory constraints for every query.

\subsection{Experiment Setup}

\noindent\textbf{Tasks and datasets.}
We use two representative tasks from StreamingBench \cite{lin2024streamingbench}, including 500 multiple-choice video--question pairs (250 per task, one question per clip): \textit{(i) Real-time visual understanding}, which focuses on the visual content in real-time, recognizing and interpreting objects, and changes within the video stream. An example query is \textit{“What text is shown on the screen?”}, and \textit{(ii) Contextual reasoning}, which focuses on the capability to comprehend the
broader context within a video stream. An example query is \textit{“How many game scores has the team scored so far?”}
To evaluate under realistic network dynamics, we adopt the 4G/5G mmWave uplink dataset \cite{ghoshal2022depth}, which includes static, walking, and driving scenarios, capturing fluctuations caused by blockage and mobility.

\noindent\textbf{Query timeline and causality.}
Video is divided into one-second segments. A query associated with segment $t$ is released when that segment closes; the controller may use frames captured up to that boundary, the static profile, and goodput observations from uploads completed by $t-1$, but no future frame or throughput sample. The response-latency clock starts at query release and ends when the backend returns the answer. Thus, the fixed one-second capture interval is not included in the reported response latency; an application that measures from the first frame of the segment should add one second to obtain information age. The main benchmark issues one outstanding query per stream. The sustained multi-device experiment in Appendix~\ref{app:devices} additionally allows consecutive requests to overlap across the capture, upload, and server pipelines while preserving per-stream timestamp order.

\begin{figure*}[!t]
\centering
{\includegraphics[width = \textwidth]{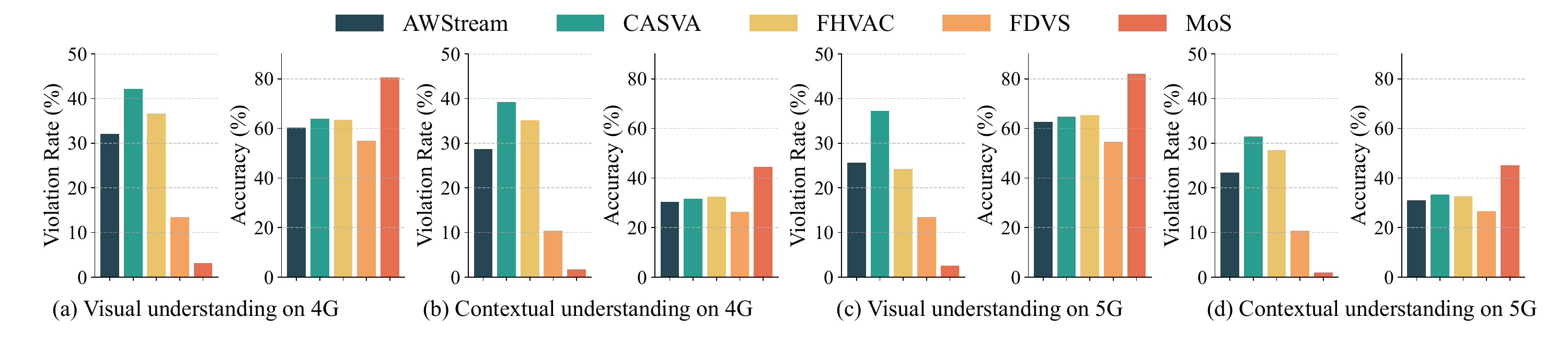}}
\vspace{-6mm}
\caption{Overall performance with the DisPider backend under different network conditions and tasks.}
\vspace{-4mm}
\label{fig:overall2}
\end{figure*}

\begin{figure}[!t]
    \centering
    \includegraphics[width=\linewidth]{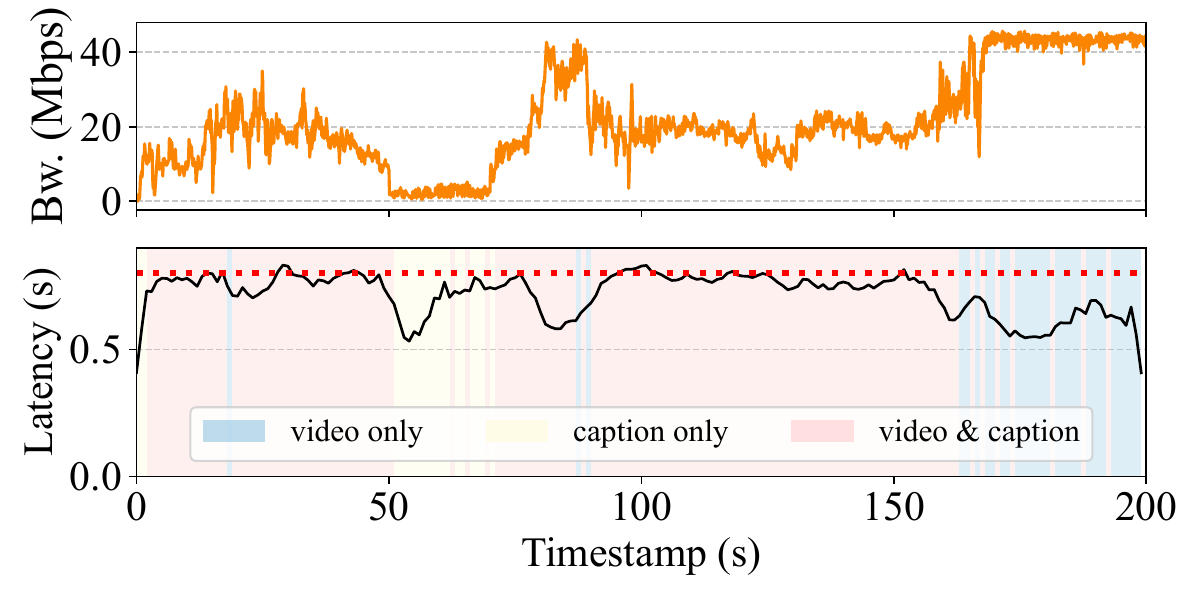}
    \vspace{-5mm}
    \caption{Runtime analysis of MoS streaming control over time.}
    \vspace{-4mm}
    \label{fig:fig8}
\end{figure}

\noindent\textbf{Baselines.}
We compare MoS against the following state-of-the-art streaming for video analysis studies:
(1) \textbf{AWStream \cite{zhang2018awstream}:} Adapts video streaming through a combination of offline and online profiling. It learns a configuration profile that models the trade-off between analytics accuracy and bandwidth usage.
(2) \textbf{CASVA \cite{zhang2022casva}:} Uses a deep reinforcement learning video streaming controller to adjust frame rate, resolution, and quantization parameter. We adopt its latency-first mode, where the edge continuously uploads the most recent video segment.
(3) \textbf{FHVAC \cite{zhang2024fhvac}:} A hybrid adaptive video streaming method that integrates rule-based logic with reinforcement learning via imitation learning.
(4) \textbf{FDVS \cite{you2024towards}:} A textual representation baseline designed for video LLM through multi-granularity information extraction and fusion. We adopt it for real-time video captioning for each segment on the edge and only transmit the converted textual captions to the server for analytics.

We also add four controls that isolate the benefit of adaptive modality mixing rather than video adaptation alone: \textbf{Always-Caption} sends an independently generated caption for every segment; \textbf{LowQ-Video+Caption} always sends the lowest video configuration and a 30-token caption; \textbf{Fixed-Mixture} uses the single video--caption pair with the best offline average accuracy subject to the 800\,ms SLO; and \textbf{Embedding-only} transmits the quantized visual embedding produced by the edge visual encoder and invokes a compatible projection layer at the server. All controls use the same segment boundaries, network traces, backend, and payload accounting as MoS. Embedding-only provides a closer representation-level comparison with recent knowledge-transmission systems such as KDC (NSDI~\textquotesingle26)~\cite{mosCommentKdc2026}; it isolates feature transmission rather than reproducing KDC\textquotesingle s full semantic-prioritization and receiver-restoration pipeline.

\noindent\textbf{Metrics.}
We evaluate the performance of MoS and the baselines using the following metrics:
\textit{(i) Analytics accuracy:} The percentage of queries for which the model selects the reference option. We parse the first unambiguous option label after normalizing case and punctuation; a missing, malformed, or multiple-option response is incorrect. Accuracy and timeliness are intentionally measured separately: a correct answer returned after the SLO remains correct for the accuracy metric and is also counted as a latency violation.
\textit{(ii) End-to-end latency:} The time from query release at the segment boundary through edge processing, transmission, and completion of cloud-side inference.
\textit{(iii) Latency violation ratio:} The percentage of queries whose end-to-end latency exceeds the target latency constraint (SLO).

\noindent\textbf{Parameter settings.}
The candidate video quality configurations, defined by resolution, frame rate, and quantization parameter, include: [360P, 3FPS, 32], [480P, 5FPS, 30], [540P, 10FPS, 28], [720P, 20FPS, 26], and [1080P, 30FPS, 23]. The candidate textual caption configurations, defined by input frame resolution and maximum output token length, include: [240P, 20], [360P, 30], [480P, 40], [540P, 50], and [720P, 60].
The profiled anchor threshold $\tau$ is set to 473. The default latency SLO is 800ms. The balancing constant $V$ in configuration selection is 200.

\subsection{Evaluation Results}
\subsubsection{Overall Performance}
\label{sec:eval-overall}

We evaluate the overall performance of MoS and the baselines under different network conditions and tasks on two backend model stacks, including VideoLLM-Online and Dspider, as shown in Fig.~\ref{fig:overall} and Fig.~\ref{fig:overall2}. Across both backends, MoS consistently achieves the highest accuracy and the lowest violation rate in all settings. On VideoLLM-Online, compared to AWStream, CASVA, FHVAC, and FDVS, MoS improves average accuracy by 40.1\%, 33.5\%, 32.8\%, and 60.4\%, respectively, while reducing average violation rates by 92.1\%, 94.2\%, 93.0\%, and 81.8\%. The results on Dspider show a similar trend: MoS improves average accuracy by 39.3\%, 32.2\%, 32.0\%, and 59.1\% over the same baselines, and reduces average violation rates by 92.2\%, 94.3\%, 92.9\%, and 82.3\%. In absolute terms, MoS achieves 80.5\% to 82.1\% accuracy on real-time visual understanding and 44.6\% to 45.1\% on contextual understanding, with only 1.1\% to 3.2\% violations, whereas the baselines exhibit substantially higher violation rates under the same network conditions.

\noindent\textbf{Additional multimodal baselines.}
Table~\ref{tab:mm-baselines} adds controls requested to separate MoS's online policy from the generic benefit of supplying text. Always-Caption has a tiny payload but pays the captioning cost on every segment. Fixed mixtures improve accuracy, but their violation rate grows when the trace leaves the bandwidth region represented by their single operating point. Embedding-only is inexpensive yet loses information at the interface to the unmodified backend. MoS obtains the strongest joint operating point by using captions selectively and retaining visual evidence when the task or network permits it.

\begin{table*}[t]
\centering
\small
\setlength{\tabcolsep}{7pt}
\renewcommand{\arraystretch}{1.08}
\caption{Additional multimodal baselines on the VideoLLM-Online backend with a Jetson Orin Nano endpoint over 4G, averaged across the two main tasks. Payload and energy include all edge-generated artifacts.}
\label{tab:mm-baselines}
\begin{tabular}{lrrrr}
\toprule
\textbf{Method} & \textbf{Accuracy (\%)} & \textbf{Violation (\%)} & \textbf{Uplink (Mbps)} & \textbf{Edge energy (J/query)} \\
\midrule
Always-Caption & 25.4 & 10.2 & 0.03 & 10.3 \\
LowQ-Video+Caption & 35.6 & 7.4 & 0.61 & 10.8 \\
Fixed-Mixture & 38.9 & 14.8 & 1.46 & 11.1 \\
Embedding-only & 28.7 & 2.8 & 0.19 & 3.2 \\
\textbf{MoS} & \textbf{40.7} & \textbf{1.9} & 0.88 & 7.6 \\
\bottomrule
\end{tabular}
\vspace{-3mm}
\end{table*}

\subsubsection{Robustness and Adaptability}
\label{sec:eval-robustness}

\noindent\textbf{Runtime analysis.}
To better understand the runtime behavior of MoS, we present a case study using a segment of the 4G LTE bandwidth trace under the real-time visual understanding task. As shown in Fig.~\ref{fig:fig8}, the top plot shows the bandwidth over time in a driving scenario, while the bottom plot illustrates the per-segment latency of MoS. The shaded regions represent the selected streaming modes: blue for video-only, yellow for text-only, and red for mixed video--text streaming. When bandwidth is sufficient (e.g., during the first 50\,s), MoS selects mixed streaming to maximize accuracy. When bandwidth becomes particularly high (e.g., after 160\,s), MoS may prefer video-only streaming because the overhead of text captioning can outweigh its accuracy benefit under the latency SLO. During bandwidth-constrained periods (e.g., around 50\,s), MoS switches to text-only or low-quality video with text, minimizing latency while preserving accuracy. This example highlights the ability of MoS to adapt its streaming mode to real-time bandwidth and maintain latency below the SLO without compromising analytics quality.

\begin{figure}[!t]
    \centering
    \includegraphics[width=\linewidth]{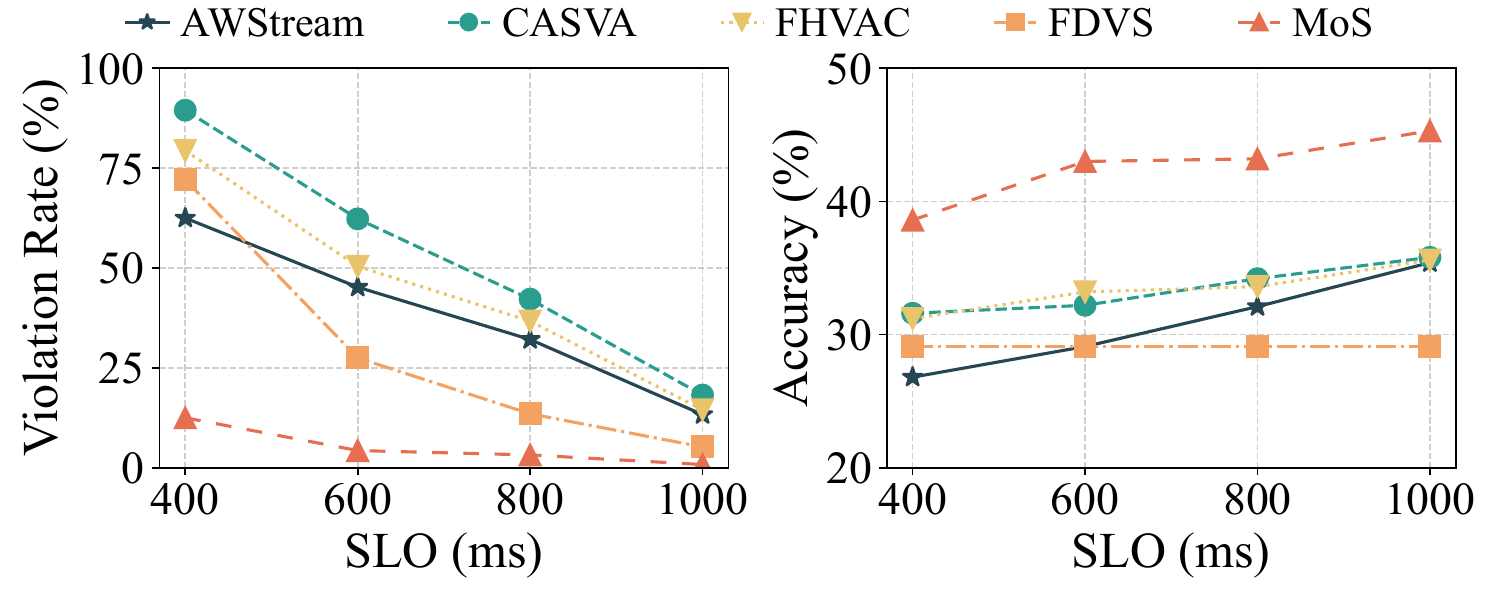}
    \vspace{-3mm}
    \caption{Accuracy and violation rate under different latency SLOs.}
    \vspace{-5mm}
    \label{fig:slo}
\end{figure}

\begin{figure*}[t]
    \centering
    \begin{minipage}[t]{0.31\linewidth}
        \centering
        \includegraphics[width=\linewidth]{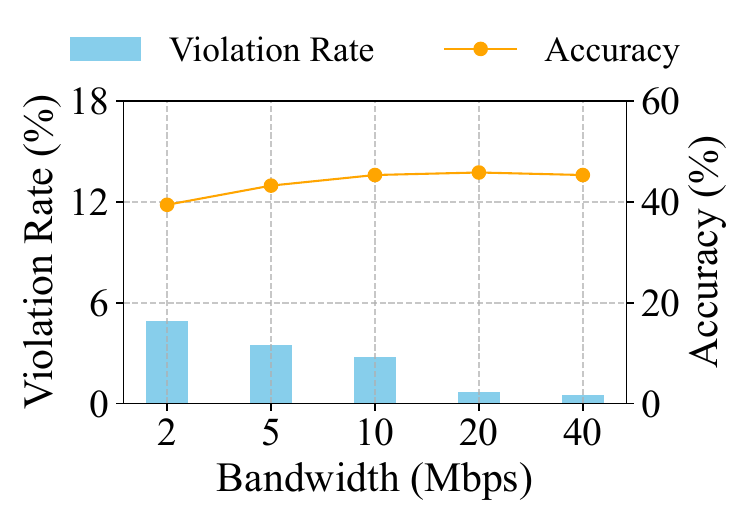}
        \caption{Accuracy and violation rate under different network settings.}
        \label{fig:diffbw}
    \end{minipage}
    \hfill
    \begin{minipage}[t]{0.31\linewidth}
        \centering
        \includegraphics[width=\linewidth]{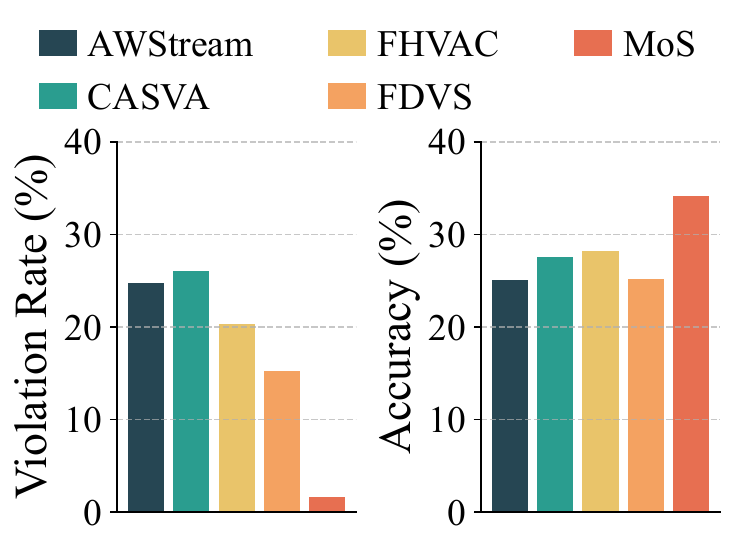}
        \caption{Performance on omni-source understanding over real-world 5G.}
        \label{fig:realwd}
    \end{minipage}
    \hfill
    \begin{minipage}[t]{0.31\linewidth}
        \centering
        \includegraphics[width=\linewidth]{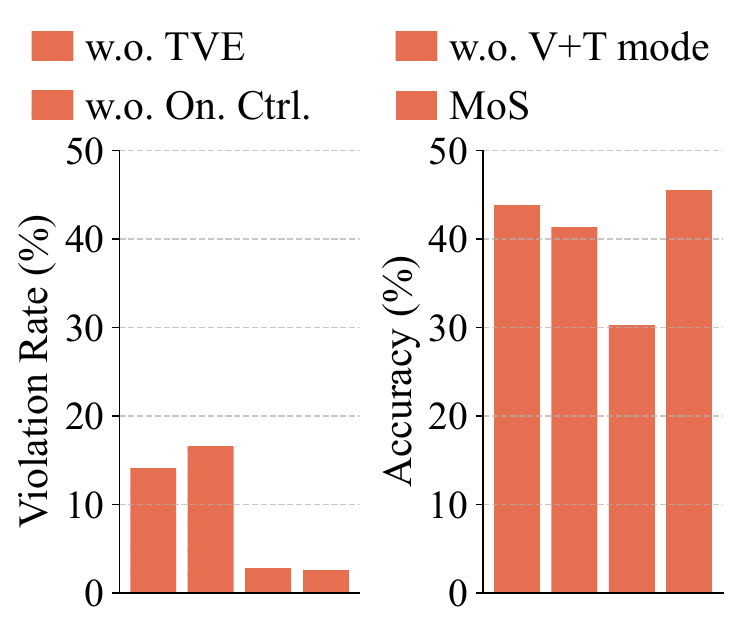}
        \caption{Ablation study of the captioning and streaming modules.}
        \label{fig:abl}
    \end{minipage}
    \vspace{-5mm}
\end{figure*}

\noindent\textbf{Performance under different latency SLOs.}
We evaluate MoS and the baselines under four latency SLOs: 400\,ms, 600\,ms, 800\,ms, and 1000\,ms. As shown in Fig.~\ref{fig:slo}, MoS consistently achieves the highest accuracy and the lowest violation rate across all SLO levels. Compared to AWStream, CASVA, FHVAC, and FDVS, MoS improves average accuracy by 37.8\%, 27.1\%, 27.3\%, and 46.1\%, respectively, while reducing violation rates by 86.4\%, 90.2\%, 88.5\%, and 82.4\%. These results demonstrate that MoS maintains strong and consistent performance across a wide range of latency requirements.

\noindent\textbf{Performance under different network settings.}
We evaluate the impact of varying bandwidth on both tasks by scaling the mean bandwidth of the network traces to 2\,Mbps, 5\,Mbps, 10\,Mbps, 20\,Mbps, and 40\,Mbps. As shown in Fig.~\ref{fig:diffbw}, MoS achieves accuracies of 39.4\%, 43.2\%, 45.3\%, 45.8\%, and 45.3\%, with corresponding violation rates of 4.9\%, 3.5\%, 2.8\%, 0.7\%, and 0.5\%, respectively. Notably, even under the most constrained bandwidth (2\,Mbps), MoS outperforms all baselines in accuracy. Across all bandwidth levels, its average latency remains within the target SLO, demonstrating robust and consistent performance under varying network conditions.

\noindent\textbf{Real-world performance on unseen data.}
We further evaluate MoS on unseen network patterns and analytics scenarios. First, we test MoS and the baselines over a real-world 5G network using our prototype, where the average bandwidth is 16.7\,Mbps. As shown in Fig.~\ref{fig:realwd}, MoS achieves the lowest average end-to-end latency, with most segments remaining within the latency SLO. In contrast, a significant portion of segments processed by the baselines exceed the latency threshold. We also evaluate MoS on an unseen task not included during offline profiling---the Omni-source Understanding task from StreamingBench. MoS continues to achieve the highest accuracy, outperforming AWStream, CASVA, FHVAC, and FDVS by 35.9\%, 23.6\%, 20.9\%, and 35.3\%, respectively. These results highlight the robustness of MoS under real-world network conditions and its generalizability to diverse analytics scenarios.

\subsubsection{Design Analysis}
\label{sec:eval-design}

\begin{figure}[!t]
    \centering
    \includegraphics[width=\linewidth]{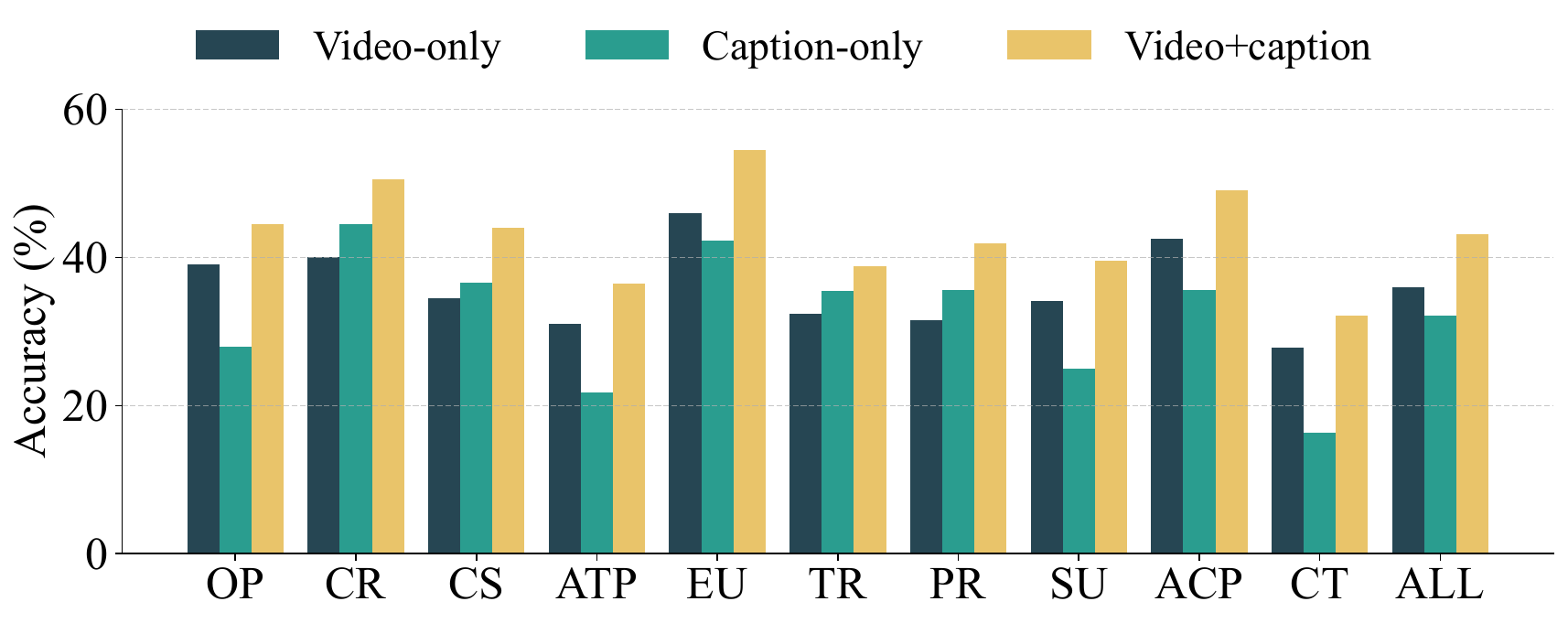}
    \vspace{-3mm}
    \caption{Performance of candidate modality representations across task categories.}
    \vspace{-5mm}
    \label{fig:modality-capacity}
\end{figure}

\noindent\textbf{Capacity analysis.}
Our results indicate that captions are an inherently lossy representation when used as the sole modality for interactive video understanding. Overall, accuracy decreases from 35.99\% with video-only input to 32.12\% with caption-only input, and the degradation is most pronounced in categories that require fine-grained visual evidence or precise aggregation over time. As shown in Fig.~\ref{fig:modality-capacity}, caption-only performs substantially worse than video-only on Object Perception (OP, 27.96\% vs. 39.07\%), Attribute Perception (ATP, 21.74\% vs. 31.05\%), Spatial Understanding (SU, 24.95\% vs. 34.16\%), Action Perception (ACP, 35.62\% vs. 42.49\%), and Counting (CT, 16.38\% vs. 27.86\%). These gaps are consistent with the failure cases observed in practice, such as identifying small patterns or logos on a person, counting how many times an event occurs across a full clip, or spotting unusual background details. In such cases, captions often omit low-saliency yet task-critical details due to aggressive compression and may hallucinate or overspecify facts, making them unreliable under an open-ended query space. Clean video+caption reaches 43.17\% overall in this diagnostic, but this is not evidence that adding text is universally beneficial: the controlled corruption study in Appendix~\ref{app:caption-error} identifies the point at which misleading text cancels the multimodal gain.

\noindent\textbf{Per-category controller decisions.}
The controller's action distribution supports the same interpretation. As summarized in Table~\ref{tab:category-selection}, it uses text-only most often for temporal/contextual questions whose salient event can be stated compactly, and retains video most often for OCR, attributes, small objects, spatial relations, and counting. The mixed mode dominates only when both a compact semantic cue and visual evidence are useful; it is not hard-coded as the default action.

\begin{table}[t]
\centering
\scriptsize
\setlength{\tabcolsep}{3pt}
\renewcommand{\arraystretch}{1.05}
\caption{Modal-selection frequency by task group over 4G.}
\label{tab:category-selection}
\begin{tabularx}{\columnwidth}{@{}Xrrr@{}}
\toprule
\textbf{Task group} & \textbf{Video (\%)} & \textbf{Text (\%)} & \textbf{V+T (\%)} \\
\midrule
Fine-grained visual/OCR & 44 & 8 & 48 \\
Temporal/contextual & 18 & 20 & 62 \\
Counting/long-range & 36 & 10 & 54 \\
All queries & 30 & 13 & 57 \\
\bottomrule
\end{tabularx}
\vspace{-3mm}
\end{table}

\begin{figure}[!t]
    \centering
    \includegraphics[width=\linewidth]{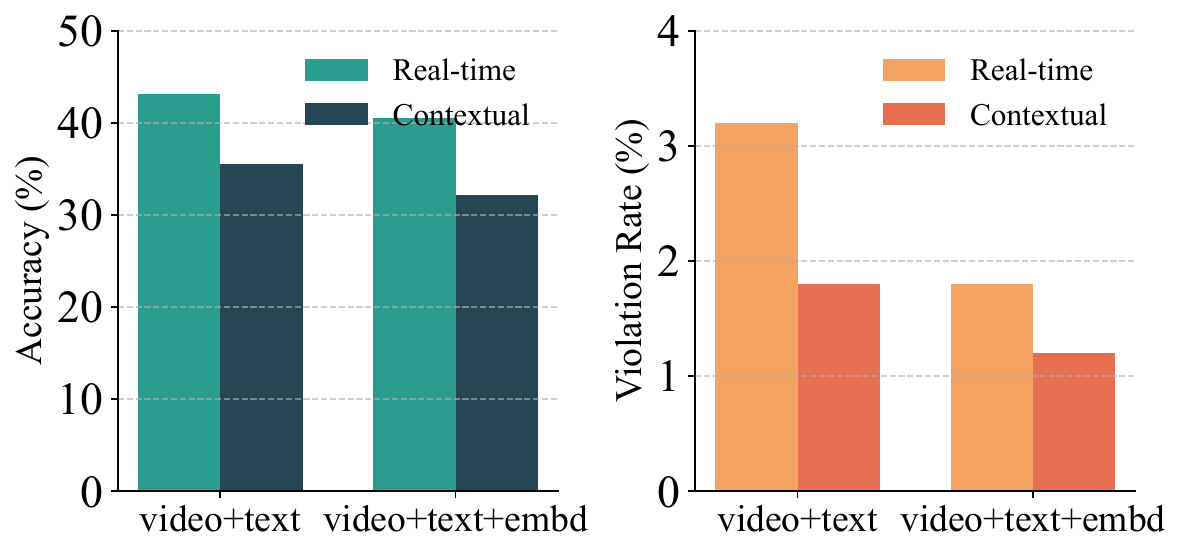}
    \vspace{-5mm}
    \caption{Performance of MoS with more candidate modalities.}
    \vspace{-4mm}
    \label{fig:more-modalities}
\end{figure}

\noindent\textbf{Performance with more modality candidates.}
We further extend MoS with a candidate stream based on visual embedding vectors, yielding a three-candidate design (video+text+embedding). Under the same 4G setting, introducing the embedding stream slightly reduces the violation rate but consistently decreases accuracy. For real-time visual understanding, MoS (video+text) achieves 43.2\% accuracy with 3.2\% violation, while MoS (video+text+embedding) drops to 40.5\% accuracy with 1.8\% violation. For contextual understanding, accuracy decreases from 35.5\% to 32.2\%, while the violation rate decreases from 1.8\% to 1.2\%. These results indicate that simply enlarging the modality candidate set does not necessarily improve end-task performance, as the embedding stream introduces additional overhead and can provide redundant or less compatible signals to the downstream model. Overall, the two-stream candidate set already captures the dominant accuracy--latency tradeoff for our workloads, and MoS benefits more from selecting an appropriate mixture than from adding more modality types.

\noindent\textbf{Ablation study.}
We evaluate the contribution of each component by comparing the following variants: \textit{(i) w.o. TVE}, which replaces the collaborative textual visual encoder with the naive static captioning approach from FDVS; \textit{(ii) w.o. online controller}, which disables online configuration adaptation and relies solely on the offline profile; and \textit{(iii) w.o. V+T mode}, which restricts each segment to a single modality and its corresponding configuration, removing the ability to combine video and text. As shown in Fig.~\ref{fig:abl}, compared to \textit{w.o. TVE}, MoS reduces the violation rate by 81.5\% while maintaining comparable accuracy. This is because static dense captioning introduces substantial latency and edge overhead without adaptive control. Compared to \textit{w.o. online controller}, MoS achieves an 84.2\% lower violation rate and 10.2\% higher accuracy. Relying only on the offline profile leads to suboptimal decisions when the input video or task distribution shifts over time. Finally, compared to \textit{w.o. V+T mode}, MoS improves accuracy by 50.7\% with a similar violation rate. This highlights the benefit of combining modalities to provide richer representations and more accurate analytics than a single modality alone.

\subsubsection{Deployment Practicality}
\label{sec:eval-practicality}

\begin{table*}[t]
\centering
\small
\setlength{\tabcolsep}{5pt}
\renewcommand{\arraystretch}{1.08}
\caption{Edge-only and edge--cloud deployment on Jetson and Pixel. Accuracy is reported separately for real-time visual (RV) and contextual reasoning (CR); violation is averaged across them at an 800\,ms SLO. }
\label{tab:videollm_online_edge}
\begin{tabular}{llrrrr}
\toprule
\textbf{Deployment} & \textbf{Endpoint accelerator} & \textbf{RV acc. (\%)} & \textbf{CR acc. (\%)} & \textbf{p50 E2E (ms)} & \textbf{Avg. viol. (\%)} \\
\midrule
Edge-only VideoLLM-Online-8B (AWQ) & Jetson GPU & 31.5 & 22.3 & $1820^\dagger$ & 100.0 \\
Edge-only MobileVLM-V2-1.7B (INT4)~\cite{chu2024mobilevlmv2} & Jetson GPU & $29.8^\dagger$ & $19.6^\dagger$ & $1430^\dagger$ & $96.8^\dagger$ \\
Edge-only MobileVLM-V2-1.7B (INT4)~\cite{chu2024mobilevlmv2} & Pixel NPU & $29.1^\dagger$ & $19.1^\dagger$ & $1180^\dagger$ & $91.4^\dagger$ \\
\midrule
\textbf{MoS edge--cloud} & \textbf{Jetson GPU} & \textbf{45.5} & \textbf{35.9} & $628^\dagger$ & \textbf{1.9} \\
\textbf{MoS edge--cloud} & \textbf{Pixel accelerator} & $\mathbf{44.7}^\dagger$ & $\mathbf{35.1}^\dagger$ & $692^\dagger$ & $4.7^\dagger$ \\
\bottomrule
\end{tabular}
\vspace{-3mm}
\end{table*}

\noindent\textbf{Comparison with edge-based models.}
We compare MoS against an edge-only deployment of VideoLLM-Online using AWQ quantization, since the full model cannot be practically deployed on the edge device without quantization. We further add MobileVLM-V2-1.7B, a mobile-oriented VLM with a lightweight projector, 4-bit weights, and 144 visual tokens~\cite{chu2024mobilevlmv2}. For a streaming comparison, the baseline receives four uniformly sampled frames from the latest segment plus a 256-token rolling textual history; it performs all reasoning locally and sends no payload. Table~\ref{tab:videollm_online_edge} separates model quality from deployability. The smaller mobile VLM reduces local latency, especially on the Pixel accelerator, but remains above the 800\,ms target for most requests and loses temporal accuracy. MoS instead keeps the endpoint model on the captioning path and offloads heavy temporal reasoning, preserving the stronger cloud backend while satisfying the SLO on both devices.

\begin{figure}[!t]
    \centering
    \includegraphics[width=\linewidth]{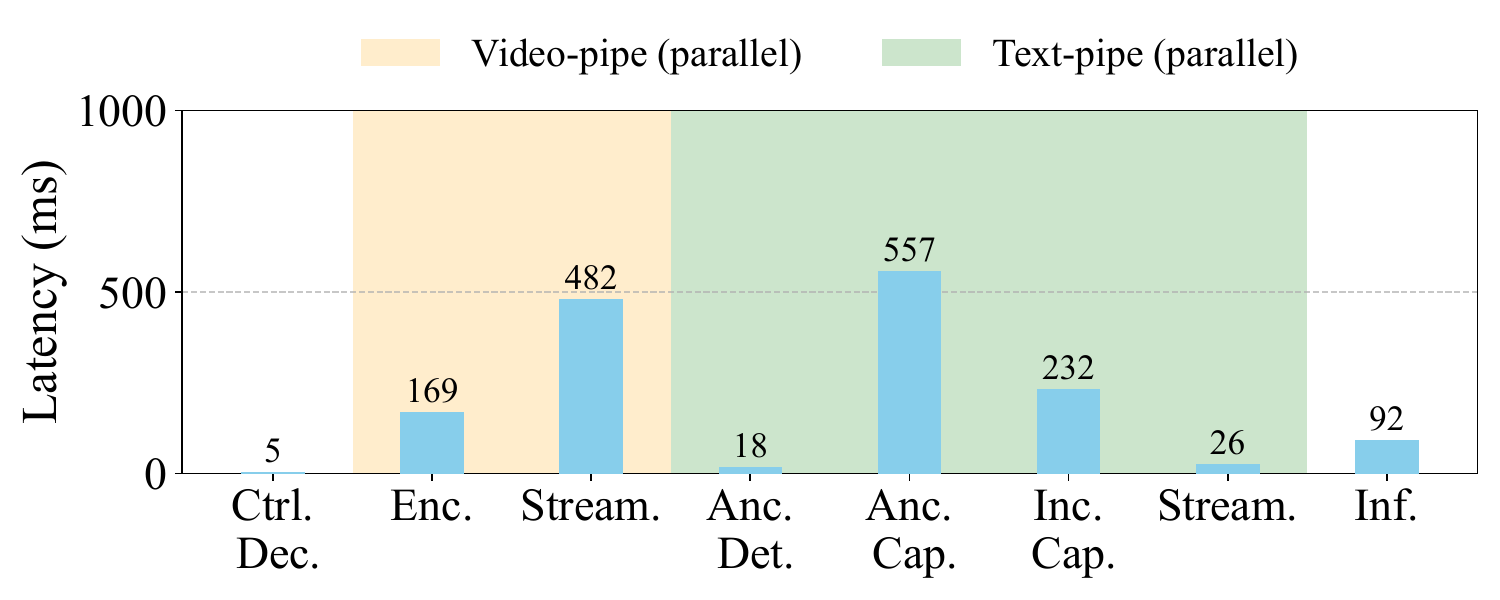}
    \vspace{-7mm}
    \caption{Component-wise overhead analysis. Abbreviations are: Ctrl. Dec., controller decision; Enc., encoding; Anc., anchor; Det., detection; Cap., caption; Inc., incremental; Inf., inference.}
    \vspace{-4mm}
    \label{fig:overhead}
\end{figure}

\noindent\textbf{Overhead analysis.}
As shown in Fig.~\ref{fig:overhead}, we provide a component-wise breakdown of the end-to-end latency in MoS to assess its system overhead. The average latency for processing a one-second video segment, from query release at the segment boundary to inference completion, remains below 800\,ms. Under 4G network conditions, the edge device spends 5\,ms on modality and configuration control. For the mixed-modality pipeline, video encoding and transmission take 169\,ms and 482\,ms, respectively. In the parallel captioning process, anchor detection and incremental captioning take 18\,ms and 232\,ms, while anchor captioning occasionally takes up to 557\,ms. On the server side, inference takes 92\,ms. Overall, processing on both sides remains within the segment duration, enabling continuous real-time operation without delay accumulation. Meanwhile, edge VRAM usage remains around 3\,GB, and CPU utilization stays below 40\%, both well within the device's capacity. This demonstrates that MoS is lightweight and suitable for deployment on common edge devices.

\section{Discussion}

MoS is most beneficial when the network is bandwidth-limited or highly dynamic, while task semantics can be partially captured by compact textual descriptions. In such cases, captions provide an efficient auxiliary signal that helps the system maintain responsiveness without fully relying on high-rate video transmission. However, the benefit of the semantic stream is task-dependent. For queries requiring fine-grained spatial details, subtle appearance cues, or precise temporal counting, captions alone may omit critical evidence and video remains indispensable. Additionally, MoS currently relies on a finite profiled configuration set and short-horizon bandwidth prediction, so its adaptation quality may degrade under severe distribution shifts or highly bursty uplink conditions. These limitations suggest future directions, including richer semantic side information beyond captions, more adaptive online profile refinement, and tighter coupling between modality scheduling and backend multimodal inference.

\noindent\textbf{Workload boundaries.}
The main benchmark uses multiple-choice queries and one outstanding request per stream. The 30-minute two-device study exercises sustained, overlapping pipelines, but it is not yet a scale-out multi-camera evaluation: dozens of simultaneous cameras would introduce server queueing and a shared-uplink allocation problem absent from the current controller. Open-ended answers would also require semantic or human grading rather than exact option matching. We therefore limit the current claims to small numbers of interactive streams and report multi-camera scheduling and open-ended evaluation as future work rather than extrapolating the present accuracy values.

\noindent\textbf{Caption failures and safe fallback.}
Object identity, OCR, counting, and spatial relations are the highest-risk caption errors because fluent text can cause a backend to overweight an incorrect semantic cue. The controlled study in Table~\ref{tab:caption-noise} explicitly tests this failure rather than assuming video+text is monotonic. A production implementation can combine cross-frame caption consistency, grounding confidence, and periodic visual audits; low confidence should remove text from the candidate pair or force video-only. This policy trades some bandwidth for a bounded failure mode and does not require changing the backend model.

\noindent\textbf{Deployment integration.}
Battery and thermal constraints enter MoS through profiled $T_{\mathrm{cap}}$ and the available configuration set, but the current objective optimizes accuracy under latency rather than energy directly. A device can remove caption configurations when a thermal or battery threshold is crossed, and a future controller can add a long-term energy queue alongside latency debt. The prototype's stream identifiers and timestamps support multiple endpoints, but privacy policy, retention, authentication, and transport security remain deployment-specific responsibilities rather than properties guaranteed by MoS.

\section{Related Work}

\subsection{Efficient Networked Video Analytics}

\noindent\textbf{Content reduction and adaptive streaming.}
Existing video analytics systems primarily reduce communication cost within the visual modality. Frame-selection systems such as Reducto~\cite{li2020reducto} and FilterForward~\cite{canel2019scaling} remove temporally redundant frames before transmission. Region-oriented systems, including DDS~\cite{du2020server}, DRE~\cite{liu2019edge}, and VaBUS~\cite{wang2022vabus}, allocate bandwidth to visually important regions. Other systems adapt encoding configurations: AWStream~\cite{zhang2018awstream}, CASVA~\cite{zhang2022casva}, and FHVAC~\cite{zhang2024fhvac} select resolutions, frame rates, and quantization levels according to content and network conditions. These approaches improve the accuracy--bandwidth tradeoff, but the transmitted representation remains video. MoS retains these video-quality knobs while additionally allowing a segment to be represented by text or by a combination of video and text.

\noindent\textbf{Online control and edge execution.}
Chameleon~\cite{jiang2018chameleon} coordinates analytics configurations across cameras, while OAVS~\cite{li2024oavs} and RIVA~\cite{li2025riva} update streaming policies as workloads change. COUPLE~\cite{bao2023couple} accelerates mobile video analytics using heterogeneous processors. These systems demonstrate the importance of jointly considering network and computing resources, but generally assume a fixed model interface and a single type of transmitted representation. MoS instead controls two asynchronous processing paths: video encoding and upload, and caption generation and upload. Its profiler accounts for the computation, communication, and backend inference costs of both paths, while its online controller selects among video-only, text-only, and video+text configurations under a common end-to-end latency constraint.

\noindent\textbf{Relationship to semantic communication.}
Semantic communication systems transmit task-relevant features instead of reconstructing the original signal, often using matched encoders and decoders or channel-aware learned representations~\cite{onsu2025semantic}. Such representations can substantially reduce bandwidth, but may be tied to a particular downstream model or task. MoS operates at a different layer: it transmits standards-compliant video and human-readable captions that can be consumed by an off-the-shelf multimodal backend. The Embedding-only baseline in our evaluation further compares MoS with feature transmission. Thus, MoS is complementary to semantic codecs, which could be introduced as additional configurations in its joint profiler.

\subsection{Video LLMs and Multimodal Representations}

\noindent\textbf{Video LLMs and online interaction.}
Video-ChatGPT~\cite{Maaz2023VideoChatGPT}, Video-LLaVA~\cite{lin2024videollava}, VideoLLaMA~\cite{zhang2023videollama}, VideoLLaMA~2~\cite{cheng2024videollama2}, and Qwen-VL~\cite{bai2023qwen} enable general-purpose video question answering, captioning, and instruction following. More recent models, including VideoLLM-Online~\cite{chen2024videollm} and DisPider~\cite{qian2025dispider}, support interaction with continuously arriving video. These studies mainly improve model architectures, training, or inference mechanisms and generally assume that visual inputs are already available to the model. MoS addresses the preceding systems problem: deciding which visual and textual information should be generated and transmitted from a resource-constrained device before the response deadline.

\noindent\textbf{Memory and visual-information reduction.}
Long-video systems such as MovieChat~\cite{song2024moviechat}, Streaming Long Video Understanding~\cite{qian2024streaming}, and LongVA~\cite{zhang2024longva} use sparse memory or long-context mechanisms to preserve temporal information. DynamicViT~\cite{rao2021dynamicvit}, MADTP~\cite{cao2024madtp}, and STC~\cite{wang2025stc} reduce inference cost through visual-token pruning or compression. These methods primarily remove information after frames have been received or encoded into visual tokens. In contrast, MoS makes a network-level decision before transmission and can replace part of the visual stream with compact textual information. Token pruning and memory compression are therefore orthogonal to MoS and can reduce the profiled backend inference cost of its candidate configurations.

{\raggedright\noindent\textbf{Caption-based representations and MoS positioning.}\par}\noindent
Dense and streaming captioning methods~\cite{krishna2017dense,yang2023vid2seq,zhou2024streaming,xue2025progress} generate temporally localized video descriptions, while FDVS~\cite{you2024towards} constructs detailed textual stories for long-video understanding. Directly applying these methods to every live segment, however, introduces continuous edge computation and may omit fine-grained visual evidence. MoS generates sparse anchor captions and motion-aware incremental updates, and treats text as a potentially lossy complement rather than a universal replacement for video. Individual techniques used by MoS—including content-change detection, caption generation, profiling, and online control—build on established ideas. Its main contribution is their modality-aware integration: a unified configuration space and cost model that schedules video, text, or both according to downstream accuracy, edge computation, transmission time, and measured latency feedback.

\section{Conclusion}
This paper presented MoS, a novel framework for real-time video LLM analytics through mixture-of-modality streaming. Motivated by the tradeoffs between visual and textual modalities, MoS enables adaptive integration of them to optimize analytics quality and system overhead. It introduces a lightweight edge-cloud collaborative captioning pipeline that incrementally summarizes visual events with minimal overhead, and a network-aware controller that adapts modality and configurations under latency constraints. Across the 500-pair StreamingBench evaluation, real 4G/5G traces, and two backend models, MoS improves analytics accuracy by up to 60.4\% and reduces violation rate by up to 94.3\% relative to the evaluated baselines. Ablation and sensitivity analysis further confirm the effectiveness of each component,
underscoring its real-world value.

\begingroup
\raggedright
\bibliographystyle{plain}
\bibliography{reference}
\endgroup

\clearpage
\appendix
\section{Controller Details}
\label{app:controller}

\subsection{Complete Configuration Selection Procedure}
\label{app:algorithm}
{\setlength{\textfloatsep}{10pt}
\begin{algorithm}[htbp]
\caption{Online Configuration Selection for Mixture-of-Modality Streaming}
\label{alg:cfg-ctrl}
\KwIn{Profile table $\mathcal P$; latency SLO $T_{\mathrm{SLO}}$;
virtual queue $Q_t$; predicted bandwidth $\hat B_t$}
\KwOut{Selected joint configuration $J_t^\star$}

$\textit{bestU}\leftarrow-\infty$,\quad
$\textit{bestJ}\leftarrow\varnothing$\;

\ForEach{$j\in\mathcal J$}{
    $\hat T_{\mathrm{vnet}}
    \leftarrow Bw_v(j)/\hat B_t$\;

    $\hat T_{\mathrm{cnet}}
    \leftarrow Bw_c(j)/\hat B_t$\;

    $\hat T_{\mathrm{tot}}
    \leftarrow
    \max\!\left(
    T_{\mathrm{cap}}(j)+\hat T_{\mathrm{cnet}},
    T_{\mathrm{enc}}(j)+\hat T_{\mathrm{vnet}}
    \right)
    +T_{\mathrm{inf}}(j)$\;

    \lIf{$\hat T_{\mathrm{tot}}>T_{\mathrm{SLO}}$}
    {\textbf{continue}}

    $U
    \leftarrow
    V\,\mathrm{Acc}(j)
    -
    Q_t\,\hat T_{\mathrm{tot}}$\;

    \If{$U>\textit{bestU}$}{
        $\textit{bestU}\leftarrow U$\;
        $\textit{bestJ}\leftarrow j$\;
    }
}

\If{$\textit{bestJ}=\varnothing$}{
    $\textit{bestJ}
    \leftarrow
    \displaystyle
    \arg\min_{j\in\mathcal J}
    \hat T_{\mathrm{tot}}(j,\hat B_t)$\;
}

\textbf{execute} $\textit{bestJ}$ and measure $T_{\mathrm{obs},t}$\;

$Q_{t+1}
\leftarrow
\max\!\left\{
0,\,
Q_t+T_{\mathrm{obs},t}-T_{\mathrm{SLO}}
\right\}$\;

\Return $\textit{bestJ}$\;
\end{algorithm}
}

\subsection{Performance Guarantee}
\label{app:guarantee}
We finally characterize the controller using standard Lyapunov
drift-plus-penalty analysis~\cite{neely2010stochastic}. The guarantee is
for long-term average latency rather than a worst-case latency bound for
every segment.

\noindent\emph{Theorem 1 (Long-term accuracy and latency guarantee).}
Assume that \textit{(i)} the network state is stationary and i.i.d. over a
bounded support, \textit{(ii)} both accuracy and end-to-end latency are
uniformly bounded over the finite configuration space, and
\textit{(iii)} for every candidate configuration, the latency estimator is
conditionally unbiased with respect to the controller's observed network
state. Further assume that there exists a stationary policy satisfying the
latency constraint with some positive slack $\epsilon>0$.

Let $Acc^\star$ denote the optimal long-term average accuracy achievable
under the same admissible configuration space and latency constraint.
Then the policy in Alg.~\ref{alg:cfg-ctrl} satisfies
\begin{equation}
\liminf_{N\rightarrow\infty}
\frac{1}{N}\sum_{t=1}^{N}
\mathbb E[\mathrm{Acc}(J_t)]
\ge
Acc^\star-\frac{\Delta}{V},
\label{eq:accuracy-bound}
\end{equation}
for some finite constant $\Delta$, while the virtual queue is mean-rate
stable and therefore
\begin{equation}
\limsup_{N\rightarrow\infty}
\frac{1}{N}\sum_{t=1}^{N}
\mathbb E\!\left[
T_{\mathrm{obs},t}-T_{\mathrm{SLO}}
\right]
\le 0.
\label{eq:latency-bound}
\end{equation}

\noindent\emph{Proof sketch.}
Define the quadratic Lyapunov function
$\mathcal L(Q_t)=\frac{1}{2}Q_t^2$. From
Eq.~\eqref{eq:queue-update}, the one-step conditional drift is bounded by
\begin{equation}
\mathbb E[
\mathcal L(Q_{t+1})-\mathcal L(Q_t)
\mid Q_t]
\le
\Delta
+
Q_t\,
\mathbb E[
T_{\mathrm{obs},t}-T_{\mathrm{SLO}}
\mid Q_t],
\end{equation}
where $\Delta$ is finite because the latency is bounded. Subtracting the
weighted utility
$V\,\mathbb E[\mathrm{Acc}(J_t)\mid Q_t]$ gives the standard
drift-plus-penalty upper bound. Since
$Q_tT_{\mathrm{SLO}}$ is independent of the selected action, minimizing
this bound is equivalent to maximizing
$V\mathrm{Acc}(j)-Q_t\hat T_{\mathrm{tot}}(j,\hat B_t)$, as in
Eq.~\eqref{eq:dpp-obj}.

Comparing this decision against an optimal stationary feasible policy,
summing the drift inequalities over time, and telescoping the Lyapunov
terms yields the $\mathcal O(1/V)$ optimality gap in
Eq.~\eqref{eq:accuracy-bound}. Under the assumed strictly feasible policy,
the virtual queue remains stable. Furthermore, from
Eq.~\eqref{eq:queue-update},
$Q_N\ge\sum_{t=1}^{N}
(T_{\mathrm{obs},t}-T_{\mathrm{SLO}})$.
Mean-rate stability therefore implies
Eq.~\eqref{eq:latency-bound}.

The parameter $V$ exposes the standard operating tradeoff of
drift-plus-penalty control: increasing $V$ brings the achieved accuracy
closer to the long-term optimum, while allowing a larger virtual queue and
hence slower correction of transient latency debt. In practice, the
predictive feasibility gate further prevents configurations that are
already expected to violate the SLO, while the virtual queue compensates
for the residual mismatch between predicted and realized latency.

\noindent\textbf{Scope of the guarantee.}
The stationary i.i.d. and conditional-unbiasedness assumptions above are sufficient conditions for the stated asymptotic bound; real 4G/5G mobility traces are temporally correlated and do not satisfy them exactly. We therefore do not interpret Theorem~1 as a per-query guarantee on those traces. The theorem explains the controller's long-run accuracy--latency tradeoff under its idealized model, whereas SLO compliance on practical traces is an empirical result. In deployment, the feasibility gate uses a causal prediction and the virtual queue is updated with realized latency, so correlated prediction errors appear as accumulated debt and make later decisions more conservative. Appendix~\ref{app:bw-error} quantifies prediction error, and Sec.~\ref{sec:eval-robustness} reports the resulting violation rate without shuffling the traces.

\section{Supplementary Experiments}
\label{app:supplementary-experiments}

\subsection{Bandwidth-Prediction Error}
\label{app:bw-error}
We replay the original traces in timestamp order and compare the one-segment-ahead prediction with realized goodput. Table~\ref{tab:bw-pred} reports absolute percentage error; the larger 5G tail reflects blockage transitions in the mmWave traces. These errors are not injected artificially and the controller receives no future samples. Despite a non-negligible tail, the realized-latency update in Eq.~\eqref{eq:queue-update} keeps the violation rate low, showing why prediction and closed-loop debt correction are both needed.

\begin{table}[t]
\centering
\small
\setlength{\tabcolsep}{6pt}
\renewcommand{\arraystretch}{1.05}
\caption{One-segment-ahead bandwidth-prediction error.}
\label{tab:bw-pred}
\begin{tabular}{lrr}
\toprule
\textbf{Trace} & \textbf{Median APE (\%)} & \textbf{90th-pct. APE (\%)} \\
\midrule
4G mobility & 14.8 & 36.9 \\
5G mmWave & 17.6 & 42.1 \\
WiFi & 7.1 & 18.4 \\
\bottomrule
\end{tabular}
\vspace{-3mm}
\end{table}

\subsection{Caption-Error Robustness}
\label{app:caption-error}
We construct four controlled error types that match observed failure modes: replacing an object identity, perturbing a count by one, deleting or corrupting OCR text, and swapping a spatial relation. We corrupt a specified fraction of caption clauses while leaving video unchanged. Table~\ref{tab:caption-noise} shows the expected trend: unfiltered video+caption degrades faster than video-only and falls below it when roughly 40\% of clauses are corrupted. A simple confidence gate---estimated from caption self-consistency across adjacent frames---falls back to video-only when confidence is low and preserves most of the clean-caption benefit. This experiment bounds the claim in Fig.~\ref{fig:modality-capacity}: captions help when sufficiently reliable, but can mislead the backend.

\begin{table}[t]
\centering
\small
\setlength{\tabcolsep}{4pt}
\renewcommand{\arraystretch}{1.05}
\caption{Accuracy under controlled caption corruption. }
\label{tab:caption-noise}
\begin{tabular}{lrrrrr}
\toprule
\textbf{Method} & \multicolumn{5}{c}{\textbf{Corrupted caption clauses (\%)}} \\
\cmidrule(lr){2-6}
& 0 & 10 & 20 & 30 & 40 \\
\midrule
Video-only & 36.0 & 36.0 & 36.0 & 36.0 & 36.0 \\
Video+caption & 43.2 & 41.6 & 39.4 & 37.1 & 34.8 \\
Video+caption w/ gate & 42.9 & 42.2 & 41.0 & 39.1 & 37.4 \\
\bottomrule
\end{tabular}
\vspace{-3mm}
\end{table}

\subsection{Sustained Two-Device Execution}
\label{app:devices}
To expose battery and thermal costs hidden by short trials, we run one Pixel stream and one Jetson stream concurrently for 30 minutes, release a query every second per stream, and retain the original chronological 4G trace. Table~\ref{tab:sustained-device} reports per-endpoint resource use; power is measured above each platform's idle level, and the temperature sensors differ (Jetson GPU die versus Pixel battery), so temperatures should not be compared across rows. Concurrency raises server GPU utilization from approximately 19\% to 34\% and modestly increases tail latency, but neither stream accumulates a segment backlog. This experiment also validates stream identifiers and timestamp-based alignment when two asynchronous modality pairs arrive at the backend.

\begin{table}[htbp]
\centering
\small
\setlength{\tabcolsep}{4pt}
\renewcommand{\arraystretch}{1.08}
\caption{Performance of a 30-minute concurrent Pixel+Jetson experiment (one query/s/device). Power is incremental over idle; energy integrates device power per answered query.}
\label{tab:sustained-device}
\begin{tabularx}{\columnwidth}{@{}>{\raggedright\arraybackslash}Xrr@{}}
\toprule
\textbf{Metric} & \textbf{Jetson Orin Nano} & \textbf{Pixel 10} \\
\midrule
Power (W) & 10.9 & 4.4 \\
Energy (J/query) & 7.8 & 3.5 \\
Peak memory (GB) & 3.1 & 5.0 \\
Final temp. ($^\circ$C) & 59.7 (GPU) & 43.1 (battery) \\
p95 E2E (ms) & 826 & 893 \\
Violation (\%) & 2.8 & 5.9 \\
\bottomrule
\end{tabularx}
\end{table}

\subsection{Energy and Long-Run Resource Cost}
\label{app:energy}
Captioning is not free even though its network payload is small. The sustained study in Table~\ref{tab:sustained-device} therefore integrates device power rather than reporting only utilization. At the planned one-query/s load, the preliminary estimate is 7.8\,J/query on Jetson and 3.5\,J/query on Pixel; on Jetson, forcing a caption on every segment raises the corresponding estimate to 10.3--11.1\,J/query for the fixed caption baselines in Table~\ref{tab:mm-baselines}, while MoS avoids some caption invocations when video-only is sufficient. The Pixel value corresponds to roughly 1.0\% of a 19.1\,Wh battery per 196 queries at the estimated active duty cycle; this conversion excludes display, cellular-radio tail energy, and other applications and should not be interpreted as full-phone battery life. We report incremental power, temperature, peak memory, and SLO violations together because any one metric alone can hide thermal throttling or queued work during continuous operation.

\end{document}